\documentclass[amsmath,amssymb,11pt,superscriptaddress,reprint, preprintnumbers, notitlepage,aps,twocolumn,nofootinbib]{revtex4-1}
\pdfoutput=1 
\usepackage[utf8]{inputenc}
\usepackage[english]{babel}
\usepackage{amsmath}
\usepackage{graphicx}
\usepackage{dcolumn}
\usepackage{pbox}
\usepackage{amssymb}
\usepackage{epsfig}
\usepackage[dvipsnames]{xcolor}
\usepackage[normalem]{ulem}
\usepackage{slashed}
\usepackage{amssymb}
\usepackage{ mathrsfs }
\usepackage{color}
\usepackage[font=small]{caption}
\usepackage[font=small]{subcaption}
\usepackage{url}
\usepackage[makeroom]{cancel}
\usepackage{tikz}
\usepackage{tikz-feynman}
\usepackage{mathptmx}
\usetikzlibrary{arrows}
\usepackage{enumitem} 
\usepackage{soul}
\usepackage{makecell}
\usepackage{pifont}
\usepackage{placeins}

\newcommand{\cmark}{\ding{51}} 
\newcommand{\xmark}{\ding{55}} 
\definecolor{MyDarkBlue}{rgb}{0.1, 0.1, 0.8} 
\definecolor{SBlue}{rgb}{0.2, 0.4, 0.7} 
\definecolor{MyLightBlue}{rgb}{0.22,0.51,0.9}
\definecolor{MyGreen}{rgb}{0.0, 0.5, 0.0}
\definecolor{BrickRed}{rgb}{0.8, 0.25, 0.33}
\RequirePackage{hyperref}
\hypersetup{colorlinks, citecolor=SBlue,linkcolor=MyDarkBlue, urlcolor=PineGreen}

\newcommand\scalemath[2]{\scalebox{#1}{\mbox{\ensuremath{\displaystyle #2}}}}

\makeatletter
\renewcommand\@makecaption[2]{%
  \par
  \vskip\abovecaptionskip
  \begingroup
   \small\rmfamily
    \begingroup
     \samepage
     \flushing
     \let\footnote\@footnotemark@gobble
     \@make@capt@title{#1}{#2}\par
    \endgroup
  \endgroup
  \vskip\belowcaptionskip
}
\makeatother

\DeclareUnicodeCharacter{2212}{-}
\begin{document}
\title{\vspace{1cm}\Large 
Neutrino Mass Induced $n$--$\overline{n}$ Oscillation
}

\author{\bf Ilja Dor\v{s}ner}
\email[E-mail:]{dorsner@phy.hr}
\affiliation{Department of Physics, Faculty of Science, University of Zagreb, HR-10000 Zagreb, Croatia}
\affiliation{Jožef Stefan Institute, Jamova 39, P.\ O.\ Box 3000, SI-1001 Ljubljana, Slovenia}

\author{\bf Svjetlana Fajfer}
\email[E-mail:]{svjetlana.fajfer@ijs.si}
\affiliation{Jožef Stefan Institute, Jamova 39, P.\ O.\ Box 3000, SI-1001 Ljubljana, Slovenia}

\author{\bf Shaikh Saad}
\email[E-mail:]{shaikh.saad@ijs.si}
\affiliation{Jožef Stefan Institute, Jamova 39, P.\ O.\ Box 3000, SI-1001 Ljubljana, Slovenia}

\begin{abstract}
The Georgi-Glashow model is the simplest possible attempt at grand unification. However, due to its particle content, the model preserves a global $U(1)_{B-L}$ symmetry, where $B$ and $L$ are baryon and lepton numbers, respectively. It thus leaves neutrinos massless just as the Standard Model does. The extensions of the Georgi-Glashow model  that break lepton number by two units, i.e., scenarios with $|\Delta L|=2$ operator(s), naturally generate potentially viable Majorana neutrino masses. Since the dynamics that yields $|\Delta L| = 2$ interactions unavoidably induces $|\Delta B| = 2$ transitions via $B\!-\!L$ breaking, these extensions  consequentially lead to intriguing processes such as neutron--antineutron ($n$--$\overline{n}$) oscillation. We investigate this intrinsic connection between neutrino mass generation and $n$--$\overline{n}$ oscillation within a number of representative extensions of the Georgi-Glashow model that can yield realistic neutrino masses and mixing parameters. These extensions include the tree-level seesaw mechanism realizations of the Type I, II, and III varieties, as well as the one-loop and two-loop radiative neutrino mass models. 
\end{abstract}

\maketitle
\section{Introduction}\label{sec:01}

The Georgi–Glashow (GG) model~\cite{Georgi:1974sy} is the simplest realization of grand unification. It achieves quark–lepton unification as well as  unification of the Standard Model (SM) interactions within $SU(5)$. However, since it predicts neutrinos to be massless, it leaves the origin of neutrino masses as one of the open questions that needs to be addressed.

We establish an intrinsic connection between Majorana neutrino mass generation in $SU(5)$ and $n$--$\overline{n}$ oscillation, both rooted in quark–lepton unification. Our work expands on a recent analysis~\cite{Dorsner:2025epy} of two simple GG model extensions of the Type II and Type III varieties that simultaneously achieve observable $n$--$\overline{n}$ oscillation~\cite{Kuzmin:1970nx,Mohapatra:1980qe,Phillips:2014fgb,Addazi:2020nlz} and accommodate realistic SM fermion mass spectrum. More specifically, we apply approach advocated in Ref.~\cite{Dorsner:2025epy} to other tree-level and radiative neutrino mass generation models to champion an unavoidable correlation between neutrino mass and $n$--$\overline{n}$ oscillation. Our work also dovetails a study~\cite{Rao:1983sd} which enumerates, in scalar extensions of the GG model, all $d=9$ operators contributing to $n$--$\overline{n}$ oscillation. 

\begin{figure}[th!]
\centering
\includegraphics[width=0.4\textwidth]{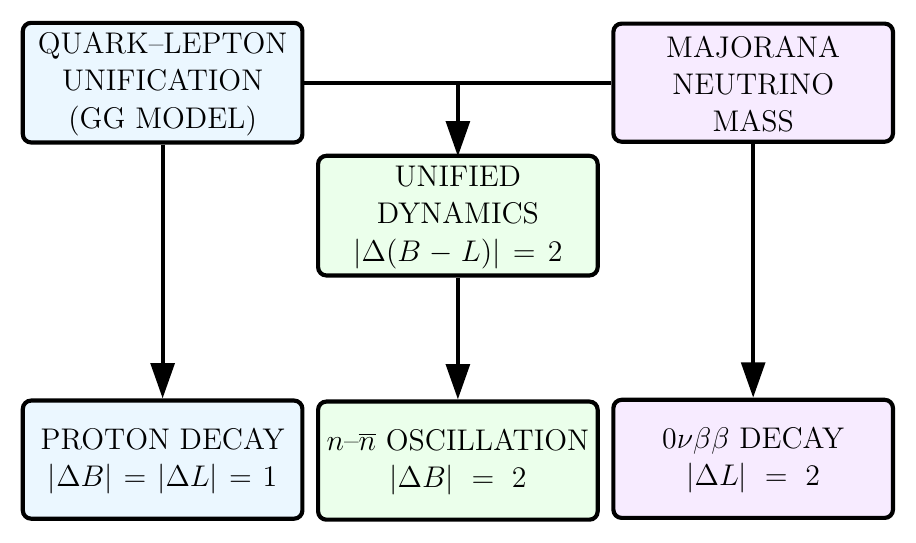}
\caption{Schematic illustrating the interplay of quark–lepton unification and neutrino mass generation with proton decay, $n$--$\overline{n}$ oscillation, and neutrinoless double beta decay ($0\nu\beta\beta$).} \label{fig:schematic}
\end{figure}

Our results are important because they imply that an observation of  baryon number–violating processes such as $n$--$\overline{n}$ oscillation could provide an independent evidence for Majorana neutrinos, beside the neutrinoless double beta decay~\cite{Dolinski:2019nrj}, if $SU(5)$ is realized in nature. Our focus on $n$--$\overline{n}$ oscillation is particularly timely given the current stringent bound from \texttt{Super-Kamiokande}~\cite{Super-Kamiokande:2020bov} and upcoming searches at \texttt{DUNE}~\cite{DUNE:2015lol} and \texttt{NNBAR}~\cite{Addazi:2020nlz} at \texttt{ESS} that are expected to significantly improve existing limits. Additional experimental handles include $|\Delta B|=1$ two-body proton decays, $|\Delta L|=2$ processes such as muon to positron conversion~\cite{Lee:2021hnx}, and $|\Delta B|=2$ transitions like di-nucleon decays~\cite{Aitken:2017wie,Girmohanta:2019cjm,Dev:2022jbf,Beneito:2025ond}. A schematic summary of these implications is shown in Fig.~\ref{fig:schematic}.

In Sec.\ \ref{sec:02}, we demonstrate that quark-lepton unification and Majorana neutrino mass generation mechanisms, in $SU(5)$, automatically yield baryon-number-violating interactions with $|\Delta B|=2$. Sec.\ \ref{sec:03} presents the construction and discussion of neutrino mass mechanisms and their implications for $n$--$\overline{n}$ oscillation. We briefly conclude in Sec.\ \ref{sec:04}.

\section{Implications of Quark-Lepton Unification and Majorana Neutrinos}\label{sec:02}
The GG model consists of the following representations:
\begin{align}
&\overline 5_{F\,i}=\left(d^c_i,L_i\right),\;\; 10_{F\,i}=\left(Q_i,u_i^c,e_i^c\right),\;\;     5_S=\left(T,H\right),\;\; 24_S,
\end{align}
where subscripts $F$ and $S$ denote fermions and scalars, respectively, and $i=1,2,3$ is a family index. Here, $Q_i$, $u^c_i$, and $d^c_i$ are the SM quark multiplets, $L_i$ and $e^c_i$ represent leptonic multiplets, $T$ stands for the color triplet, while $H$ is the SM Higgs doublet. 
Given the above particle content, $U(1)_{B-L}$ remains an accidental global symmetry of the model \cite{Weinberg:1979sa, Weinberg:1980bf, Wilczek:1979et, Wilczek:1979hc, Weldon:1980gi, Rao:1983sd}. To demonstrate this, let us look at the Yukawa part of the Lagrangian
\begin{align}
    \mathcal{L}_Y &= 10_{F} \overline 5_{F} 5_S^* + 10_{F} 10_{F} 5_S 
 \rightarrow
    \left(Q L+u^c d^c\right)T^* \nonumber \\
    &+ \left(Q Q+u^c e^c\right)T  
    +\left(e^c L+Q d^c\right)H^* + \left(Q u^c\right)H.\label{eq:main}  
\end{align}
One can trivially check that interactions of Eq.\ \eqref{eq:main} preserve a global $U(1)_{B-L}$ symmetry with
\begin{align}
Q_{+1/3},\;\; u_{-1/3}^c,\;\; d^c_{-1/3},\;\; L_{-1},\;\; e^c_{+1},\;\; T_{-2/3},\;\; H_{0},    
\end{align}
where the subscripts represent the $U(1)_{B-L}$ charges. To correct the charged fermion mass relations in the GG model, one can minimally extend it by introducing a single $45_S$~\cite{Georgi:1979df}. But, even with the additional Yukawa interactions, $U(1)_{B-L}$ remains unbroken~\cite{Rao:1983sd}. Note that $U(1)_{B-L}$ conservation allows for two-body proton decays such as $p\to e^+\pi^0$, where both initial and final states have $B-L=+1$.

Introduction of Majorana neutrino masses, which requires $|\Delta L|=2$, automatically breaks $U(1)_{B-L}$ by two units, thus inducing $|\Delta B|=2$ processes. We can illustrate this via an effective field theory approach~\cite{deGouvea:2014lva}. Specifically, an analogue of the $d=5$ Weinberg operator within the GG framework yields
\begin{align}
\scalemath{0.96}
{     
\mathcal{L}_5=    \frac{C_{ij}}{\Lambda} \overline 5_{F\,i} \overline 5_{F\,j}  5_S 5_S  \rightarrow \underbrace{ \left( \frac{C^{\nu \nu}_{ij}}{\Lambda_1} \underbrace{ L_iL_jHH }_{\Delta L=+2}  + \frac{C^{d d}_{ij}}{\Lambda_2} \underbrace{  d^c_id^c_jTT }_{\Delta B=-2} \right)}_{\Delta(B-L)=- 2}   
}, \label{eq:weinberg}
\end{align}
where $\Lambda$ denotes a new physics scale and $C_{ij}$'s are the associated Wilson coefficients. We accommodate model dependence of relevant scales and couplings in the $SU(5)$ broken phase through introduction of $\Lambda_{1,2}$ and $C^{\nu \nu,d d}_{ij}$. The dynamics generating Majorana neutrino masses  clearly induces  $|\Delta B|=2$ processes. Namely, the last term in Eq.~\eqref{eq:weinberg} generates $n$--$\overline{n}$ oscillation through the following $d=9$ operator:
\begin{align}
\label{eq:d9operator}
 \frac{C^{dd}_{11}}{\Lambda_2}  d_1^cd_1^cTT \to  \frac{C^{dd}_{11} C^{ud}_{11}}{\Lambda_2 \Lambda^4_3}  d_1^cd_1^c \left(u_1^cd_1^c\right) \left(u_1^cd_1^c\right).
\end{align}

To better understand advocated connection between quark-lepton unification and Majorana neutrino mass generation within the framework of renormalizable $SU(5)$ models, let us consider the following two examples:
\begin{itemize}
\item \textbf{Type I seesaw.} If one introduces a Majorana fermion $1_F=\nu^c(1,1,0)$, the Yukawa sector
\begin{align}
\mathcal{L}_Y&\supset y_{i} \, \overline 5_{F\,i} 1_{F} 5_S  
 \rightarrow y^\nu_{i} \, L_i \nu^c H + y^{d}_{i} \, d_i^c \nu^c T, 
 \label{eq:typeone}
\end{align}
requires $\nu^c$ to carry $B-L$ charge of $+1$. Therefore, its Majorana mass term
\begin{align}
\mathcal{L}_Y&\supset    \underbrace{M^{\nu} \, \nu^c \nu^c }_{\Delta(B-L)=- 2},
\end{align}
breaks $B-L$ by two units.

\item \textbf{Type II seesaw.} This case requires $15_S=\Delta_{1}(1,3,1)+\Delta_{3}(3,2,1/6)+\Delta_{6}(\overline{6},1,-2/3)$. The Yukawa interaction reads
\begin{align}
\mathcal{L}_Y & \supset y_{ij} \,\overline 5_{F\,i} \overline 5_{F\,j} 15_S\nonumber\\
&\rightarrow y^\nu_{ij}\, L_i L_j\Delta_{1}  + y^{Ld}_{ij}\,L_i d_j^c \Delta_{3} + y^{dd}_{ij}\, d_i^c d_j^c \Delta_{6},
\end{align}
which assigns $B-L$ charges of $-2$, $+4/3$, and $+2/3$ to $\Delta_1$, $\Delta_3$, and $\Delta_6$, respectively. 
The scalar potential includes 
\begin{align}
V&\supset \mu\, 5_S 5_S 15^*_S \rightarrow \underbrace{ \overline\mu \, HH \Delta_{1}^* + \hat\mu\, H T \Delta_{3}^* + \widetilde\mu\, TT\Delta_{6}^*  }_{\Delta(B-L)=- 2},
\end{align}
where each term breaks $B-L$ by two units.
\end{itemize}

This analysis  yields the same conclusion in any other renormalizable $SU(5)$ model of Majorana neutrino mass.

\begin{table*}[t]
\centering
\small
\begin{tabular}{c|c|c|c|c|c}
\hline
\textbf{MODEL} & \textbf{CONTENT} & $\nu$ \textbf{MASS} & \textbf{CLASS} & \textbf{TOPOLOGY} & $n$--$\overline{n}$\; \textbf{MEDIATORS}  \\
\hline
\makecell{Georgi–Glashow\\ (GG) \cite{Georgi:1974sy}} & $\overline{5}_{F\,i} + 10_{F\,i} + 5_S + 24_S$ & -- & -- & -- & -- \\ \hline

\makecell{Type~I seesaw\\\cite{Minkowski:1977sc,Yanagida:1979as,Glashow:1979nm,Gell-Mann:1979vob,Mohapatra:1979ia}} &  $\mathrm{GG}+1_F+1_F$ & tree level & \textbf{I} & \textbf{B} &
\makecell[l]{
$S^{u^cd^c}_{1}=5^*_S,\;\; F^{d^c}_1=1_F$ \\
$S_{1}=(\overline 3,1,1/3),\;\; F_1=(1,1,0)$
} \\  \hline

\makecell{Type~II seesaw\\\cite{Magg:1980ut,Schechter:1980gr,Lazarides:1980nt,Mohapatra:1980yp}} & \makecell{$\mathrm{GG}+15_S$\\\cite{Ma:1998dn,Dorsner:2005fq,Dorsner:2005ii,Dorsner:2006hw,Dorsner:2007fy,Antusch:2022afk,Calibbi:2022wko,Antusch:2023mqe,Kaladharan:2024bop,Dev:2025sox,Dorsner:2025epy}} & tree level & \textbf{II} & \textbf{A} &
\makecell[l]{
$S^{u^cd^c}_{1}=5^*_S,\;\; S^{d^cd^c}_2=15_S$ \\
$S_{1}=(\overline 3,1,1/3),\;\; S_2=(6,1,-2/3)$
}  \\  \hline

\makecell{Type~II seesaw\\ variant} & \makecell{ $\mathrm{GG}+15_S+45_S$\\ \cite{Dorsner:2025epy}} & tree level & \textbf{II} & \textbf{A} &
\makecell[l]{
$S^{u^cd^c}_{1}=45^*_S,\;\; S^{d^cd^c}_2=15_S$ \\
$S_{1}=(6,1,1/3),\;\; S_2=(6,1,-2/3)$
}  \\  \hline

\makecell{Type~III seesaw\\\cite{Foot:1988aq}} & \makecell{$\mathrm{GG}+24_F+24_F$ \\ \cite{Ma:1998dn,Bajc:2006ia,Dorsner:2006fx,FileviezPerez:2007bcw,Antusch:2023kli}} & tree level & \textbf{I} & \textbf{B} &
\makecell[l]{
$S^{u^cd^c}_{1}=5^*_S,\;\; F^{d^c}_1=24_F$ \\
$S_{1}=(\overline 3,1,1/3),\;\; F_1=(1,1,0), (8,1,0)$
}  \\  \hline

\makecell{Type~III seesaw\\ variant} & \makecell{$\mathrm{GG}+24_F+45_S$ \\ \cite{Dorsner:2025epy}} & tree level & \textbf{I} & \textbf{B} &
\makecell[l]{
$S^{u^cd^c}_{1}=45^*_S,\;\; F^{d^c}_1=24_F$ \\
$S_{1}=(6,1,1/3),\;\; F_1=(8,1,0)$
}  \\  \hline

\makecell{BNT\\\cite{Babu:2009aq}} & \makecell{$\mathrm{GG}+15_F+35_S$ \\ \cite{Dorsner:2019vgf,Dorsner:2021qwg,Antusch:2023jok,Dorsner:2024jiy}} & \makecell{tree \& \\ one-loop} & \textbf{III} & \textbf{C} &
\makecell[l]{
$S^{u^cd^c}_{1}=5_S, \; S_2=35_S$, \; $F^{d^c}_1=15_F$ \\
$S_{1}=(3,1,-1/3),\;\; S_2=(\overline{10},1,1)$,\\  $F_1=(6,1,-2/3)$
}  \\  \hline

\makecell{Zee\\\cite{Zee:1980ai}} & \makecell{$\mathrm{GG}+10_S+45_S$ \\ \cite{Perez:2016qbo,Klein:2019jgb,Dorsner:2025rzj}} & one-loop & \textbf{II} & \textbf{AI} &
\makecell[l]{
$S^{u^cd^c}_{1}=45^*_S,\;\; S^{d^cd^c}_2=10_S$ \\
$S_{1}=(6,1,1/3),\;\; S_2=(\overline{3},1,-2/3)$
}  \\  \hline

\makecell{Zee-Babu\\\cite{Cheng:1980qt,Babu:1988ki}} & \makecell{$\mathrm{GG}+10_S+50_S$ \\ \cite{Babu:2024jdw,Saad:2019vjo}} & two-loop & \textbf{II} & \textbf{AII} &
\makecell[l]{
$S^{d^cd^c}_{1}=10_S,\;\; S^{u^cu^c}_2=50_S$ \\
$S_{1}=(\overline{3},1,-2/3),\;\; S_2=(6,1,4/3)$
}  \\ \hline

\makecell{Zee-Babu\\ variant} & \makecell{$\mathrm{GG}+15_S+50_S$ \\ } & \makecell{tree \& \\two-loop} & \textbf{II} & \textbf{A} &
\makecell[l]{
$S^{d^cd^c}_{1}=15_S,\;\; S^{u^cu^c}_2=50_S$ \\
$S_{1}=(6,1,-2/3),\;\; S_2=(6,1,4/3)$
\\
\hline
$S^{\overline{Q}\overline{Q}}_{1}=50^*_S,\;\; S^{d^cd^c}_2=15_S$ \\
$S_{1}=(6,3,1/3),\;\; S_2=(6,1,-2/3)$
}  \\

\hline
\end{tabular}

\caption{Summary of the viable neutrino mass models and the associated particle content within the GG framework of quark-lepton unification. The third column indicates the order of the neutrino mass generation. The fields listed in the sixth column correspond to those providing the leading $n$--$\overline{n}$ contribution, with their $SU(3)\times SU(2)\times U(1)$ transformation properties explicitly shown in the second line. Superscript $q$ in $F^{q}$ represents the type of quark $q$ that fermion $F$ interacts with, while $S^{q\hat q}$ represents a scalar with an interaction of the $S^{q\hat q}q\hat q$ type. The $n$--$\overline{n}$ diagram topologies are denoted with {\bf A}, {\bf AI}, {\bf AII}, {\bf B}, and {\bf C}, where all these topologies can be found in lower panels of Figs.\ \ref{fig:fig01}, \ref{fig:fig02}, \ref{fig:fig05}, \ref{fig:fig03}, and \ref{fig:fig04}.    }
\label{tab:models}
\end{table*}

\section{Models}\label{sec:03}
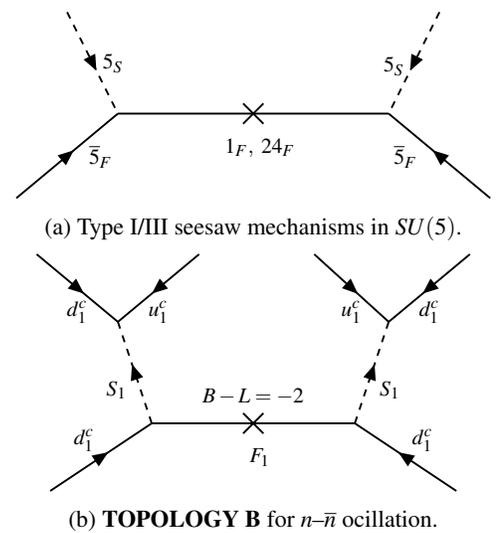
\begin{figure}[b!]
\centering

\begin{subfigure}[t]{0.45\textwidth}
\centering
\scalebox{0.9}{
\begin{tikzpicture}
\begin{feynman}
  \vertex (a3) at (-3.5,-1.25);
  \vertex (b3) at (3.5,-1.25);
  \coordinate (v1) at (-2.75,1.5);
  \coordinate (v2) at (-2,0);
  \coordinate (v3) at (2.75,1.5);
  \coordinate (v4) at (2,0);
  \coordinate (mid) at (0,0);

  \diagram* {
    (v1) -- [charged scalar, thick] (v2),
    (a3) -- [fermion, thick] (v2),
    (v3) -- [charged scalar, thick] (v4),
    (b3) -- [fermion, thick] (v4),
    (v2) -- [plain, thick] (v4),
  };

  \node at ($(v1)!0.5!(v2) + (0.3,-0.01)$) {\(5_S\)};
  \node at ($(v3)!0.5!(v4) + (-0.3,-0.05)$) {\(5_S\)};
  \node at ($(a3)!0.5!(v2) + (0.5,-0.01)$) {\(\overline 5_F\)};
   \node at ($(b3)!0.5!(v4) + (-0.5,-0.01)$) {\(\overline 5_F\)};
  \draw[thick] (mid) ++(-0.15,0.15) -- ++(0.3,-0.3);
  \draw[thick] (mid) ++(-0.15,-0.15) -- ++(0.3,0.3);
  \node at (0.1,-0.5) {\(1_F,\;24_F\)};
\end{feynman}
\end{tikzpicture}
}
\caption{Type I/III seesaw mechanisms in $SU(5)$.}
\label{fig:typeI_III}
\end{subfigure}
\hfill

\begin{subfigure}[t]{0.45\textwidth}
\centering
\scalebox{0.9}{
\begin{tikzpicture}
  \begin{feynman}
    \vertex (a1) at (-0.8,2.5);
    \vertex (a2) at (-3.2,2.5);
    \vertex (b1) at (0.9,2.5);
    \vertex (b2) at (3.2,2.5);
    \vertex (a3) at (-3,-1);
    \vertex (b3) at (3,-1);
    \coordinate (v1) at (-2,1.5);
    \coordinate (v2) at (-1.5,0);
    \coordinate (v3) at (2,1.5);
    \coordinate (v4) at (1.5,0);
    \coordinate (mid) at ($(v2)!0.5!(v4)$);

    \diagram* {
      (a1) -- [fermion, thick] (v1),
      (v2) -- [charged scalar, thick, edge label={\(S_1\)}] (v1),
      (a2) -- [fermion, thick] (v1),
      (a3) -- [fermion, thick] (v2),
      (b1) -- [fermion, thick] (v3),
      (v4) -- [charged scalar, thick, edge label'={\(S_1\)}] (v3),
      (b2) -- [fermion, thick] (v3),
      (b3) -- [fermion, thick] (v4),
      (v2) -- [plain, thick] (v4),
    };
    \draw[thick] (mid) ++(-0.15,0.15) -- ++(0.3,-0.3);
    \draw[thick] (mid) ++(-0.15,-0.15) -- ++(0.3,0.3);
    \node at ($(mid)+(0,0.4)$) {\(B-L=-2\)};

    \node at (0.1,-0.5) {\(F_1\)};
    \node at ($(a3)!0.5!(v1) + (0,-0.5)$) {\(d^c_1\)};
    \node at ($(b3)!0.5!(v3) + (0,-0.5)$) {\(d^c_1\)};
    \node at ($(a1)!0.5!(v1) + (0,-0.35)$) {\(u^c_1\)};
    \node at ($(b1)!0.5!(v3) + (0,-0.35)$) {\(u^c_1\)};
    \node at ($(a2)!0.5!(v1) + (0,-0.35)$) {\(d^c_1\)};
    \node at ($(b2)!0.5!(v3) + (0,-0.35)$) {\(d^c_1\)};
  \end{feynman}
\end{tikzpicture}
}
\caption{\textbf{TOPOLOGY B} for $n$--$\overline{n}$ ocillation.}
\label{fig:typeII}
\end{subfigure}

\caption{Type I/III seesaws (upper panel) and the associated topology, i.e., \textbf{TOPOLOGY B}, behind $n$--$\overline{n}$ oscillation (lower panel).}
\label{fig:fig01}
\end{figure}

We focus on minimal renormalizable extensions of the GG model that generate phenomenologically viable Majorana neutrino masses and, consequentially, yield $n$--$\overline{n}$ oscillation signatures. Here, minimality refers both to the number of employed representations beyond those of the GG model as well as the dimensionality of these additional representations. More precisely, we do not consider extensions with more than two additional representations, where their dimensionality should not exceed fifty. The GG model extensions relevant for our work are listed in Table~\ref{tab:models}. The associated neutrino mass diagrams are presented in the upper panels of Fig.\ \ref{fig:fig01} through Fig.\ \ref{fig:fig04}, whereas the corresponding $n$--$\overline{n}$ topologies are given in the lower panels of Fig.\ \ref{fig:fig01} through Fig.\ \ref{fig:fig04}. 

The second column of Table~\ref{tab:models} lists the particle content of each scenario, the third column gives the neutrino mass generation order, and the fourth and fifth columns specify the class and topology of $n$--$\overline{n}$ oscillation. The last column lists the mediators of the $d=9$ operators responsible for $n$--$\overline{n}$. The \textbf{CLASS} indicates the $U(1)_{B-L}$ breaking mechanism. \textbf{CLASS I} requires at least one Majorana fermion, \textbf{CLASS II} requires scalar(s), while \textbf{CLASS III} requires both a Dirac fermion and a scalar. If the model generates multiple contributions to $n$--$\overline{n}$ oscillation, we lists in Table~\ref{tab:models}  particle content yielding the dominant transition rate. 

\textbf{Type I seesaw model}: 
One needs at least two right-handed neutrinos to generate viable neutrino mass parameters in the Type I seesaw. The interplay between a scalar $(\overline{3},1,1/3) \in 5^*_S$ and a fermion $1_F$ produces the $n$--$\overline{n}$ diagram of \textbf{TOPOLOGY B}, as shown in the lower panel of Fig.~\ref{fig:fig01}. The fermion featured in \textbf{TOPOLOGY B} is always of Majorana nature, and  the $|\Delta(B-L)|=2$ breaking arises from its Majorana mass. Scalar $(\overline{3},1,1/3) \in 5^*_S$ can mediate proton decay and its contribution towards $n$--$\overline{n}$ signal is expected to be suppressed.

\begin{figure}[b!]
\centering

\begin{subfigure}[t]{0.45\textwidth}
\centering
\scalebox{0.9}{ 
\begin{tikzpicture}
\begin{feynman}
  \vertex (a3) at (-2,0);
  \vertex (b3) at (2,0);
  \vertex (mid) at (0,0);
  \vertex (top) at (0,2);
  \vertex (ul) at (-1,3);
  \vertex (ur) at (1,3);

  \diagram* {
    (a3) -- [fermion, thick] (mid),
    (b3) -- [fermion, thick] (mid),
    (top) -- [charged scalar, dashed, thick] (mid),
    (ul)  -- [charged scalar, dashed, thick] (top),
    (ur)  -- [charged scalar, dashed, thick] (top),
  };

  \node at ($(a3)!0.5!(mid) + (0,0.3)$) {\(\overline{5}_F\)};
  \node at ($(b3)!0.5!(mid) + (0,0.3)$) {\(\overline{5}_F\)};
  \node at ($(mid)!0.5!(top) + (0.35,0)$) {\(15_S\)};
  \node at ($(ul)!0.5!(top) + (-0.7,0.3)$) {\(5_S\)};
  \node at ($(ur)!0.5!(top) + (0.7,0.3)$) {\(5_S\)};
\end{feynman}
\end{tikzpicture}
}
\caption{Type II seesaw mechanism in $SU(5)$.}
\label{fig:typeII_seesaw}
\end{subfigure}
\hfill

\begin{subfigure}[t]{0.45\textwidth}
\centering
\scalebox{0.9}{ 
\begin{tikzpicture}
  \begin{scope}[rotate=180]
  \begin{feynman}
    \vertex (a) at (0,0);
    \vertex (x1) at (-1.5, 0);
    \vertex (x2) at (1.5, 0);
    \vertex (x3) at (0,-1.5);
    \vertex (f1) at (-3, 1.6);
    \vertex (f2) at (-3, -1.6);
    \vertex (f3) at (3, 1.6);
    \vertex (f4) at (3, -1.6);
    \vertex (f5) at (-1.5, -3);
    \vertex (f6) at (1.5, -3);

    \diagram* {
      (a) -- [charged scalar, thick,  edge label=\(S_1\)] (x1),
      (a) -- [charged scalar, thick,  edge label'=\(S_1\)] (x2),
      (a) -- [charged scalar, thick,  edge label=\(S_2\)] (x3),

      (f1) -- [fermion, thick, edge label'=\(u_1^c\)] (x1),
      (f2) -- [fermion, thick, edge label=\(d_1^c\)] (x1),

      (f3) -- [fermion, thick, edge label=\(u_1^c\)] (x2),
      (f4) -- [fermion, thick, edge label'=\(d_1^c\)] (x2),

      (f5) -- [fermion, thick, edge label'=\(d_1^c\)] (x3),
      (f6) -- [fermion, thick, edge label=\(d_1^c\)] (x3),
    };
\node at ($(a)+(0,0.4)$) {\(B\!-\!L=-2\)};
  \end{feynman}
  \end{scope}
\end{tikzpicture}%
}
\caption{\textbf{TOPOLOGY A} for $n$--$\overline{n}$ oscillation.}
\label{fig:zee_babu}
\end{subfigure}

\caption{Type II seesaw mechanism (upper panel) and the associated topology, i.e., \textbf{TOPOLOGY A}, behind $n$--$\overline{n}$ oscillation (lower panel).}
\label{fig:fig02}
\end{figure}
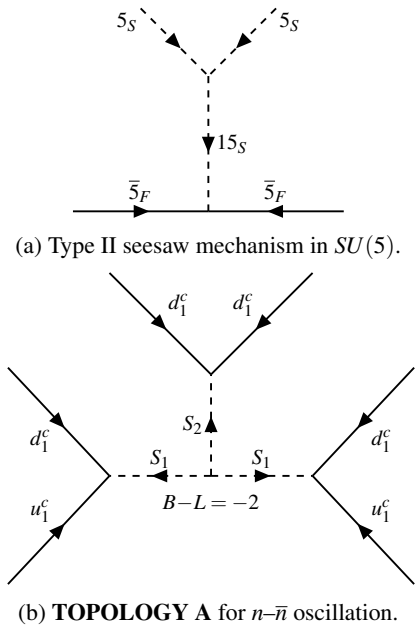

\textbf{Type II seesaw model}:  
The Type~II seesaw corresponds to \textbf{CLASS II}, with the $n$--$\overline{n}$ transition via \textbf{TOPOLOGY A} that is mediated by $(\overline{3},1,1/3) \in 5^*_S$ and $(6,1,-2/3) \in 15_S$. The neutrino mass and $n$--$\overline{n}$ diagrams are featured in Fig.\ \ref{fig:typeII_seesaw} and Fig.\ \ref{fig:zee_babu}, respectively.

\textbf{Type II seesaw variant}:
This interesting variant employs two color sextets, $S_1 = (6,1,1/3) \in 45^*_S$ and $(6,1,-2/3) \in 15_S$, to mediate $n$--$\overline{n}$ oscillation. Since these sextets do not induce proton decay, the oscillation time can be within the observable range~\cite{Dorsner:2025epy}. The neutrino mass and $n$--$\overline{n}$ diagrams appear in the upper and lower panels of Fig.~\ref{fig:fig02}, respectively.

\textbf{Type III seesaw model}: 
One requires two copies of $24_F$ to generate viable neutrino mass parameters even though a single $24_F$ already contains two states acting as right-handed neutrinos. The $B-L$ symmetry is broken by the Majorana mass of $24_F$. The neutrino mass and $n$--$\overline{n}$ diagrams are shown in the upper and lower panels of Fig.~\ref{fig:fig01}, respectively.

\textbf{Type III seesaw variant}: 
One can replace one $24_F$ with $45_S$ to achieve consistency with neutrino oscillation data. The $n$--$\overline{n}$ mediators, $(6,1,1/3) \in 45^*_S$ and $(8,1,0) \in 24_F$, do not induce proton decay and can enhance the oscillation rate~\cite{Dorsner:2025epy}. Mediators of $n$--$\overline{n}$ oscillation can also be $(\overline{3},3,1/3) \in 45^*_S$ and $(1,3,0) \in 24_F$, but the relevant  couplings of the scalar multiplet are antisymmetric in flavor space and cannot generate the leading-order contribution. We accordingly omit this particular scenario from Table~\ref{tab:models}.

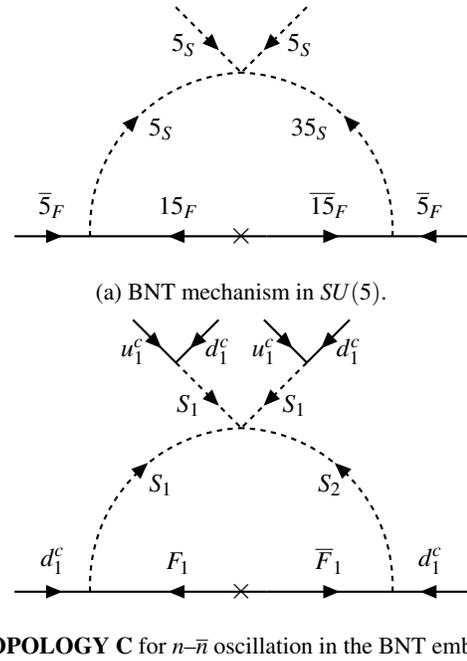
\begin{figure}[t!]
\centering

\begin{subfigure}[t]{0.45\textwidth}
\centering
\scalebox{0.67}{
\begin{tikzpicture}[inner sep=5pt]
\tikzfeynmanset{arrow size=2.1pt}
\tikzset{
  every edge/.append style={line width=1.43pt},
  every arrow/.append style={scale=1.43},
}
\begin{feynman}
\vertex(a1);
\vertex[right=2cm of a1](a1p);
\vertex[right=11cm of a1](a5);
\vertex[right=3.5cm of a1](a2); 
\vertex[right=9.5cm of a1](a3); 
\vertex[right=7cm of a1](a4);     
\vertex(a6) at ($(a2)!0.5!(a3) + (0,3.25cm)$); 
\coordinate (phiL) at ($(a6)+(-1.3cm,1.3cm)$);
\coordinate (phiR) at ($(a6)+(1.3cm,1.3cm)$);
\vertex at ($(a2)!0.5!(a3)$) (mid1);

\diagram* {
(a1p)--[fermion,  line width=1.2pt, edge label={\tikz[baseline]{\node[font=\Large]{\(\overline{5}_F\)};}}](a2), 
(a2) --[scalar, quarter left,  line width=1.2pt,  pos=0.5, with arrow=0.5](a6),  
(a3) --[scalar, quarter right,  line width=1.2pt, pos=0.5, with arrow=0.5](a6), 
(a4) --[fermion,  line width=1.2pt, edge label'={\tikz[baseline]{\node[font=\Large]{\(15_F\)};}}](a2),
(a4) --[fermion,  line width=1.2pt, edge label={\tikz[baseline]{\node[font=\Large]{\(\overline{15}_{F}\)};}}](a3),
(a5)--[fermion,  line width=1.2pt, edge label'={\tikz[baseline]{\node[font=\Large]{\(\overline{5}_{F}\)};}}](a3), 
(phiL) --[scalar,   line width=1.2pt, with arrow=0.55] (a6),
(phiR) -- [scalar,  line width=1.2pt, with arrow=0.55] (a6),
};
\node[scale=2] at (mid1) {$\times$};
\node at ($(phiL)!0.5!(a6) + (1.8,-0.1)$) {\tikz[baseline]{\node[font=\Large]{\(5_S\)};}};
\node at ($(phiR)!0.5!(a6) + (-1.8,-0.1)$) {\tikz[baseline]{\node[font=\Large]{\(5_S\)};}};
\node at ($(a2)!0.5!(a6) + (-0.1,0.5)$) {\tikz[baseline]{\node[font=\Large]{\(5_S\)};}};
\node at ($(a3)!0.5!(a6) + (-0.15,0.5)$) {\tikz[baseline]{\node[font=\Large]{\(35_S\)};}};
\end{feynman}
\end{tikzpicture}
}
\caption{BNT mechanism in $SU(5)$.}
\label{fig:nu_mass_SU5}
\end{subfigure}
\hfill

\begin{subfigure}[t]{0.45\textwidth}
\centering
\scalebox{0.67}{
\begin{tikzpicture}[inner sep=5pt]
\tikzfeynmanset{arrow size=2.1pt}
\begin{feynman}
\vertex(a1);
\vertex[right=2cm of a1](a1p);
\vertex[right=11cm of a1](a5);
\vertex[right=3.5cm of a1](a2); 
\vertex[right=9.5cm of a1](a3); 
\vertex[right=7cm of a1](a4);     
\vertex[above=3.25cm of a4](a6);  

\vertex(a6) at ($(a2)!0.5!(a3) + (0,3.25cm)$); 
\coordinate (phiL) at ($(a6)+(-1.3cm,1.3cm)$);
\coordinate (phiR) at ($(a6)+(1.3cm,1.3cm)$);
\vertex at ($(a2)!0.5!(a3)$) (mid1);

\coordinate (uL) at ($(phiL)+(-0.85cm,0.85cm)$);
\coordinate (dL) at ($(phiL)+(0.85cm,0.85cm)$);
\coordinate (uR) at ($(phiR)+(-0.85cm,0.85cm)$);
\coordinate (dR) at ($(phiR)+(0.85cm,0.85cm)$);

\diagram* {
(a1p)--[fermion, line width=1.2pt, edge label={\tikz[baseline]{\node[font=\Large]{\(d^c_1\)};}}](a2), 
(a2) --[scalar, quarter left, line width=1.2pt, pos=0.55, with arrow=0.55](a6),  
(a3) --[scalar, quarter right, line width=1.2pt, pos=0.45, with arrow=0.55](a6), 
(a4) --[fermion, line width=1.2pt, edge label'={\tikz[baseline]{\node[font=\Large]{\(F_1\)};}}](a2),
(a4) --[fermion, line width=1.2pt, edge label={\tikz[baseline]{\node[font=\Large]{\(\overline F_1\)};}}](a3),
(a5)--[fermion, line width=1.2pt, edge label'={\tikz[baseline]{\node[font=\Large]{\(d^c_1\)};}}](a3), 

(phiL) --[scalar, line width=1.2pt, with arrow=0.55] (a6),
(phiR) --[scalar, line width=1.2pt, with arrow=0.55] (a6),
(uL) --[fermion, line width=1.2pt] (phiL),
(dL) --[fermion, line width=1.2pt] (phiL),
(uR) --[fermion, line width=1.2pt] (phiR),
(dR) --[fermion, line width=1.2pt] (phiR)
};
\node[scale=1.9] at (mid1) {$\times$};
\node at ($(a2)!0.5!(a6) + (-0.1,0.5)$) {\tikz[baseline]{\node[font=\Large]{\(S_1\)};}};
\node at ($(a3)!0.5!(a6) + (0.25,0.5)$) {\tikz[baseline]{\node[font=\Large]{\(S_2\)};}};
\node at ($(phiL)!0.5!(a6) + (-0.4,-0.2)$) {\tikz[baseline]{\node[font=\Large]{\(S_1\)};}};
\node at ($(phiR)!0.5!(a6) + (0.4,-0.2)$) {\tikz[baseline]{\node[font=\Large]{\(S_1\)};}};

\node at ($(uL)!0.5!(phiL) + (-0.43,-0.2)$) {\tikz[baseline]{\node[font=\Large]{\(u^c_1\)};}};
\node at ($(dL)!0.5!(phiL) + (0.4,-0.2)$) {\tikz[baseline]{\node[font=\Large]{\(d^c_1\)};}};
\node at ($(uR)!0.5!(phiR) + (-0.43,-0.2)$) {\tikz[baseline]{\node[font=\Large]{\(u^c_1\)};}};
\node at ($(dR)!0.5!(phiR) + (0.4,-0.2)$) {\tikz[baseline]{\node[font=\Large]{\(d^c_1\)};}};
\end{feynman}
\end{tikzpicture}
}
\caption{\textbf{TOPOLOGY C} for $n$--$\overline{n}$ oscillation in the BNT embedding.}
\label{fig:nnbar_SU5}
\end{subfigure}

\caption{Babu–Nandi–Tavartkiladze (BNT) mechanism (upper panel) and associated $n$--$\overline{n}$ oscillation diagram (lower panel).}
\label{fig:fig05}
\end{figure}

\textbf{BNT model}:
The Babu--Nandi--Tavartkiladze (BNT) model is the only model in Table~\ref{tab:models} employing a Dirac fermion to generate both neutrino masses and $n$--$\overline{n}$ oscillation. The upper panel of Fig.~\ref{fig:fig05} shows the dominant contribution to the neutrino mass. A tree-level term also exists but it is negligible~\cite{Dorsner:2019vgf}. The $n$--$\overline{n}$ mediators  are $(3,1,-1/3) \in 5_S$, $(\overline{10},1,1) \in 35_S$, and $(6,1,-2/3) \in 15_F$, corresponding to \textbf{TOPOLOGY C}.  

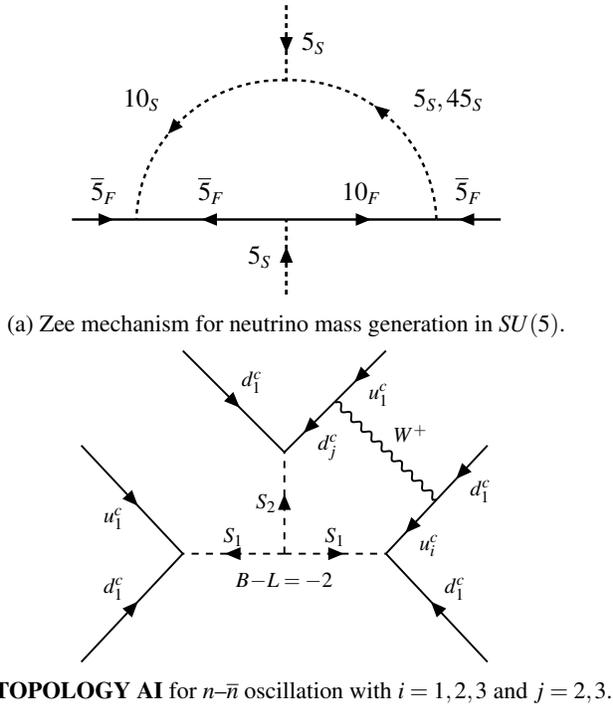
\begin{figure}[b!]
\centering

\begin{subfigure}[t]{0.45\textwidth}
\centering
\scalebox{0.57}{ 
\begin{tikzpicture}[inner sep=5pt]
\tikzfeynmanset{arrow size=2.3pt}
\begin{feynman}
\vertex(a1);
\vertex[right=2cm of a1](a1p);
\vertex[right=12cm of a1](a5);
\vertex[right=3.5cm of a1](a2); 
\vertex[below=0.5mm of a2](label1);
\vertex[right=9.5cm of a1](a3); 
\vertex[right=7cm of a1](a4);     
\vertex[below=0.75mm of a3](label2);
\vertex[above=3.25cm of a4](a6);  
\vertex[above=1mm of a6](label3);
\vertex[right=1.75cm of a2](b1);  
\vertex[right=1.75cm of a4](b2);    
\vertex[right=1.75cm of b2](a3); 
\vertex[above=1.75cm of a6](b8);  

\vertex at ($(a2)!0.5!(a3)$) (mid1); 
\vertex[below=1.75cm of mid1](c1);

\diagram* {
(a1p) -- [fermion, ultra thick, 
          edge label={\tikz[baseline]{\node[font=\LARGE]{$\overline{5}_F$};}}] (a2), 
(a6) -- [scalar, quarter right, ultra thick, 
         edge label'={\tikz[baseline]{\node[font=\LARGE]{$10_S$};}}, 
         pos=0.55, with arrow=0.55] (a2),  
(a3) -- [scalar, quarter right, ultra thick, 
         edge label'={\tikz[baseline]{\node[font=\LARGE]{{$5_S,45_S$}};}}, 
         pos=0.45, with arrow=0.55] (a6), 
(a4) -- [fermion, ultra thick, 
         edge label'={\tikz[baseline]{\node[font=\LARGE]{$\overline{5}_F$};}}] (a2),
(a4) -- [fermion, ultra thick, 
         edge label={\tikz[baseline]{\node[font=\LARGE]{$10_F$};}}] (a3),
(a5) -- [fermion, ultra thick, 
         edge label'={\tikz[baseline]{\node[font=\LARGE]{$\overline{5}_F$};}}] (a3),
(b8) --[charged scalar, ultra thick](a6),
(c1) --[charged scalar, ultra thick](mid1),
(b8) -- [charged scalar, ultra thick, 
         edge label={\tikz[baseline]{\node[font=\LARGE]{$5_S$};}}] (a6),
(c1) -- [charged scalar, ultra thick, 
         edge label={\tikz[baseline]{\node[font=\LARGE]{$5_S$};}}] (mid1)
};
\end{feynman}
\end{tikzpicture}
}
\caption{Zee mechanism for neutrino mass generation in $SU(5)$.}
\label{fig:scalar_loop}
\end{subfigure}
\hfill

\begin{subfigure}[t]{0.45\textwidth}
\centering
\scalebox{0.9}{ 
\begin{tikzpicture}
  \begin{scope}[rotate=180]
    \begin{feynman}
      \vertex (a) at (0,0);
      \vertex (x1) at (-1.5, 0); 
      \vertex (x2) at (1.5, 0);   
      \vertex (x3) at (0,-1.5);   
      \vertex (f1) at (-3, 1.6);
      \vertex (f2) at (-3, -1.6);
      \vertex (f3) at (3, 1.6);
      \vertex (f4) at (3, -1.6);
      \vertex (f5) at (-1.5, -3);
      \vertex (f6) at (1.5, -3);
      \vertex (vWstart) at ($(f5)!0.5!(x3)$);
      \vertex (vWend)   at ($(f2)!0.5!(x1)$);

\coordinate (midx) at ($(x1)!0.5!(x2)$);
\node at ($(midx)+(0,0.4)$) {\(B\!-\!L=-2\)};
      \diagram* {
        (a) -- [charged scalar, thick, edge label=\(S_1\)] (x1),
        (a) -- [charged scalar, thick, edge label'=\(S_1\)] (x2),
        (a) -- [charged scalar, thick, edge label=\(S_2\)] (x3),

        (f1) -- [fermion, thick, edge label'=\(d_1^c\)] (x1),
        (f2) -- [fermion, thick, edge label=\(d_1^c\)] (vWend),
        (vWend) -- [fermion, thick, edge label=\(u^c_i\)] (x1),

        (f3) -- [fermion, thick, edge label=\(d_1^c\)] (x2),
        (f4) -- [fermion, thick, edge label'=\(u_1^c\)] (x2),

        (f5) -- [fermion, thick, edge label=\(u_1^c\)] (vWstart),
        (vWstart) -- [fermion, thick, edge label=\(d^c_j\)] (x3),

        (f6) -- [fermion, thick, edge label=\(d^c_1\)] (x3),

        (vWstart) -- [boson, edge label=\(W^+\), thick] (vWend),
      };
    \end{feynman}
  \end{scope}
\end{tikzpicture}%
}
\caption{\textbf{TOPOLOGY AI} for $n$--$\overline{n}$ oscillation with $i=1,2,3$ and $j=2,3$.}
\label{fig:w_loop}
\end{subfigure}

\caption{Zee mechanism (upper panel) and the associated topology, i.e., \textbf{TOPOLOGY AI}, behind $n$--$\overline{n}$ oscillation (lower panel).}
\label{fig:fig03}
\end{figure}

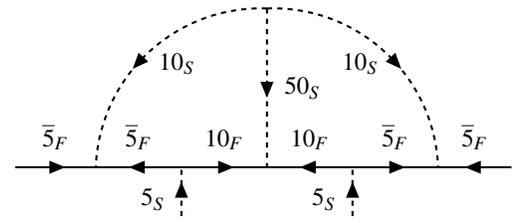
\begin{figure}[t!]
\centering

\begin{subfigure}[t]{0.45\textwidth}
\centering
\scalebox{0.65}{
\begin{tikzpicture}[inner sep=5pt]
\tikzfeynmanset{arrow size=2.1pt}
\begin{feynman}
\vertex(a1);
\vertex[right=1.85cm of a1](a1p);
\vertex[right=12cm of a1](a5);
\vertex[right=3.5cm of a1](a2); 
\vertex[below=0.5mm of a2](label1);
\vertex[right=9.5cm of a1](a3);  
\vertex[right=7cm of a1](a4);     
\vertex[below=0.75mm of a3](label2);
\vertex[above=3.25cm of a4](a6);  
\vertex[above=1mm of a6](label3);
\vertex[right=1.75cm of a2](b1);  
\vertex[below=1.cm of b1](b6);
\vertex[right=1.75cm of a4](b2);    
\vertex[right=1.75cm of b2](a3); 
\vertex[below=1.cm of b2](b7);

\diagram* {
(a1p) -- [fermion, line width=1.2pt, 
          edge label={\tikz[baseline]{\node[font=\Large]{$\overline{5}_F$};}}] (a2), 
(a6) -- [scalar, quarter right, line width=1.2pt, 
         pos=0.55, with arrow=0.55] (a2),  
(a6) -- [scalar, quarter left, line width=1.2pt, 
         pos=0.45, with arrow=0.55] (a3), 
(b1) -- [fermion, line width=1.2pt, 
         edge label={\tikz[baseline]{\node[font=\Large]{$10_F$};}}] (a4),
(b2) -- [fermion, line width=1.2pt, 
         edge label'={\tikz[baseline]{\node[font=\Large]{$10_F$};}}] (a4),
(b2) -- [fermion, line width=1.2pt, 
         edge label={\tikz[baseline]{\node[font=\Large]{$\overline{5}_F$};}}] (a3),
(a5) -- [fermion, line width=1.2pt, 
         edge label'={\tikz[baseline]{\node[font=\Large]{$\overline{5}_F$};}}] (a3),
(a6) -- [scalar, line width=1.2pt, with arrow=0.5, 
         edge label={\tikz[baseline]{\node[font=\Large]{$50_S$};}}] (a4),
(b1) -- [fermion, line width=1.2pt, 
         edge label'={\tikz[baseline]{\node[font=\Large]{$\overline{5}_F$};}}] (a2),
(b6) -- [charged scalar, line width=1.2pt, 
         edge label={\tikz[baseline]{\node[font=\Large, yshift=-20pt]{$5_S$};}}] (b1),
(b7) -- [charged scalar, line width=1.2pt, 
         edge label={\tikz[baseline]{\node[font=\Large, yshift=-20pt]{$5_S$};}}] (b2),
};
\node at ($(a2)!0.5!(a6) + (-0.1,0.5)$) {\tikz[baseline]{\node[font=\Large]{\(10_S\)};}};
\node at ($(a3)!0.5!(a6) + (0.15,0.5)$) {\tikz[baseline]{\node[font=\Large]{\(10_S\)};}};
\end{feynman}
\end{tikzpicture}
}
\caption{Zee-Babu mechanism for neutrino mass generation.}
\label{fig:scalar_loop2}
\end{subfigure}
\hfill

\begin{subfigure}[t]{0.45\textwidth}
\centering
\scalebox{0.9}{ 
\begin{tikzpicture}
  \begin{scope}[rotate=180]
    \begin{feynman}
      \vertex (a) at (0,0);
      \vertex (x1) at (-1.5, 0); 
      \vertex (x2) at (1.5, 0);   
      \vertex (x3) at (0,-1.5);   
      \vertex (f1) at (-3, 1.6);
      \vertex (f2) at (-3, -1.6);
      \vertex (f3) at (3, 1.6);
      \vertex (f4) at (3, -1.6);
      \vertex (f5) at (-1.5, -3);
      \vertex (f6) at (1.5, -3);
      \vertex (vWstartR) at ($(f5)!0.5!(x3)$);
      \vertex (vWendR)   at ($(f2)!0.5!(x1)$);

      \vertex (vWstartL) at ($(f6)!0.5!(x3)$);
      \vertex (vWendL)   at ($(f4)!0.5!(x2)$);
\coordinate (midx) at ($(x1)!0.5!(x2)$);
\node at ($(midx)+(0,0.4)$) {\(B\!-\!L=-2\)};

      \diagram* {
        (a) -- [charged scalar, thick, edge label=\(S_1\)] (x1),
        (a) -- [charged scalar, thick, edge label'=\(S_1\)] (x2),
        (a) -- [charged scalar, thick, edge label=\(S_2\)] (x3),

        (f2) -- [fermion, thick, edge label=\(u_1^c\)] (vWendR),
        (vWendR) -- [fermion, thick, edge label=\(d^c_l\)] (x1),

        (f5) -- [fermion, thick, edge label=\(d_1^c\)] (vWstartR),
        (vWstartR) -- [fermion, thick, edge label=\(u^c_i\)] (x3),

        (f6) -- [fermion, thick, edge label'=\(d_1^c\)] (vWstartL),
        (vWstartL) -- [fermion, thick, edge label=\(u^c_j\)] (x3),

        (f4) -- [fermion, thick, edge label'=\(u_1^c\)] (vWendL),
        (vWendL) -- [fermion, thick, edge label=\(d^c_k\)] (x2),

        (f1) -- [fermion, thick, edge label'=\(d_1^c\)] (x1),
        (f3) -- [fermion, thick, edge label=\(d_1^c\)] (x2),

        (vWstartR) -- [boson, edge label=\(W^+\), thick] (vWendR),
        (vWstartL) -- [boson, edge label=\(W^+\), thick] (vWendL),
      };
    \end{feynman}
  \end{scope}
\end{tikzpicture}%
}
\caption{\textbf{TOPOLOGY AII} for $n$--$\overline{n}$ oscillation with $i,j=1,2,3$ and $k,l=2,3$.}
\label{fig:w_loop2}
\end{subfigure}

\caption{Zee-Babu mechanism in $SU(5)$ (upper panel) and the associated topology behind $n$--$\overline{n}$ oscillation (lower panel).}
\label{fig:fig04}
\end{figure}

\textbf{Zee model}: 
The $\mathrm{GG}+10_S$ model can generate neutrino masses and $n$--$\overline{n}$ process but cannot accommodate realistic neutrino oscillation data. The  $\mathrm{GG}+10_S+45_S$ extension resolves this. Here, the $n$--$\overline{n}$ mediators are $S_1=(6,1,1/3) \in 45^*_S$ and $(\overline{3},1,-2/3) \in 10_S$. Due to antisymmetric nature of Yukawa couplings of $10_S$, a $W$-boson exchange is required for the transition. The neutrino mass and $n$--$\overline{n}$ diagrams are shown in the upper and lower panels of Fig.~\ref{fig:fig03}, respectively, with the latter corresponding to \textbf{TOPOLOGY AI}.

\begin{table*}
\centering
\small
\begin{tabular}{c| c| c| c| c| c| c|c|c}
\hline
MODEL & $C^{\nu\nu}_{\tau\tau}$  & $\Lambda_1$ & $C^{dd}_{11} C^{ud}_{11}$ & $\Lambda_2\Lambda_3^4$   & $\mathrm{(\texttt{NNBAR})} \geq \Lambda_2\Lambda_3^4 \geq \mathrm{(\texttt{SK})}$  & $F_k/S_k$ &\texttt{LHC}&PD\\
\hline
Type I  & $(y^\nu)^2$ & $M_{F_1} \simeq 6\times  10^{11}$ & $(y^{dd})^2 (y^{ud})^2$ & $M_{F_1}M_{S_1}^4$    & $3 \geq   M_{S_1} \geq 1.8$ & -&\xmark&\xmark\\ 
\hline
\makecell{Type II\\Type II variant}  & $y^\nu$ & $\frac{M^2_{S_3}}{\overline\mu} \simeq 6\times  10^{11}$  & $y^{dd} (y^{ud})^2$ & $\frac{M^4_{S_1}M^2_{S_2}}{\widetilde \mu}$    & $4.2\times 10^3 \geq M_{S_{2}}^{1/4}M_{S_1} \geq 2.6\times 10^3$  & $S_3: (1,3,1)$&\makecell{\cmark\\\cmark}&\makecell{\xmark\\\cmark}\\
\hline
\makecell{Type III\\Type III variant}  & $(y^\nu)^2$ & $M_{F_1} \simeq 6\times  10^{11}$  & $(y^{dd})^2 (y^{ud})^2$ & $M_{F_{2}}M_{S_1}^4$   & $2.6\times 10^3 \geq M_{F_{2}}^{1/4}M_{S_1} \geq 1.6\times 10^3$  & $F_2:(8,1,0)$ &\makecell{\cmark\\\cmark}&\makecell{\xmark\\\cmark}\\
\hline
BNT & $\frac{y^\nu \hat y^\nu}{16\pi^2}$ & $\frac{M^2_{S_3}}{M_{F_2}} \simeq 6\times  10^{11}$ & $\frac{y^{d} \hat y^{d} (y^{ud})^2}{16\pi^2}$ & $\frac{M^4_{S_1}M^2_{S_2}}{M_{F_1}}$    & $1.3\times 10^3\geq \frac{M_{S_1}M^{1/2}_{S_2}}{M_{F_1}^{1/4}} \geq 8.3\times 10^2$ & \makecell{$F_2:(1,3,1)$ \\ $S_3:(1,4,-3/2)$} &\cmark&\xmark\\
\hline
Zee & $\frac{y^\nu \hat y^\nu y_\tau}{16\pi^2}$ & $\frac{M^2_{S_3}}{\overline\mu} \simeq 4\times  10^{7}$ & $\frac{g^2_2V_{ub}V_{td}}{16\pi^2} \frac{m_bm_t}{m^2_W}  y^{ds} \left(y^{ud}\right)^2$ & $\frac{M^4_{S_1}M^2_{S_2}}{\widetilde \mu}$    & $125 \geq M_{S_{1}}M_{S_2}^{1/4} \geq 77$  & $S_3:(1,1,1)$&\cmark&\cmark\\
\hline
\makecell{Zee-Babu\\Zee-Babu variant} & $\frac{y^\nu \hat y^\nu y^2_\mu}{256\pi^4}$ & $\frac{M^2_{S_3}}{\overline\mu} \simeq 140$ &  \makecell{$\frac{g^4_2V_{ub}^2V_{td}^2}{256\pi^4} \frac{m_b^2m_t^2}{m^4_W} \left(y^{ds}\right)^2 y^{uu}$\\$y^{uu} (y^{dd})^2$} & $\frac{M^4_{S_1}M^2_{S_2}}{\widetilde \mu}$   & \makecell{$3.7 \geq M_{S_{2}}^{1/4}M_{S_1} \geq  2.3$\\$4.2\times 10^3 \geq M_{S_{2}}^{1/4}M_{S_1} \geq 2.6\times 10^3$}  &  \makecell{$S_3:(1,1,1)$\\$S_3:(1,3,1)$} &\makecell{\xmark\\\cmark}&\makecell{\cmark\\\cmark}\\
\hline
\end{tabular}
\caption{The mass scales and couplings relevant for neutrino masses and $n$--$\overline{n}$ process, where all the mass parameters are given in TeV units. $V_{ij}$ are the quark mixing parameters and $g_2$ is the $SU(2)$ gauge coupling of the SM. PD denotes viability with respect to proton decay limits. }\label{tab:EFT}
\end{table*}

\textbf{Zee-Babu model}:
In the Zee–Babu model, both neutrino masses and $n$--$\overline{n}$ oscillation arise at two loops, as shown in Fig.~\ref{fig:fig04}. The $n$--$\overline{n}$ mediators are $S_1=(\overline{3},1,-2/3) \in 10_S$ and $(6,1,4/3) \in 50_S$, corresponding to \textbf{TOPOLOGY AII}. Due to antisymmetric nature of Yukawa coupling of $10_S$ and the fact that scalar cubic interaction involves two $10_S$ propagators, two $W$-boson exchanges are required to generate $n$--$\overline{n}$ transition.

\textbf{Zee-Babu variant}: 
We note that $10_S$ of the minimal Zee–Babu model can be replaced by $15_S$, which would enhance the $n$--$\overline{n}$ rate due to introduction of symmetric Yukawa couplings of $15_S$ to the SM fermions. There would thus be no need for two $W$-boson exchanges to generate $n$--$\overline{n}$ oscillation. In this particular variant a singly charged component within $(1,3,1)\in 15_S$ replaces $(1,1,1)\in 10_S$ in the Zee–Babu two-loop neutrino mass diagram of Fig.\ \ref{fig:scalar_loop2}.

This variant also adds a tree-level contribution to neutrino masses, which might be subdominant in parts of parameter space. For example, this could be trivially achieved by assigning a tiny vacuum expectation value to scalar $(1,3,1)\in 15_S$. If that is the case, the leading $n$--$\overline{n}$ contributions would be via \textbf{TOPOLOGY A} of Fig.\ \ref{fig:zee_babu} and could arise entirely via color sextets, with two mediator options: $(6,1,-2/3) \in 15_S$ and $(6,1,4/3) \in 50_S$, or $(6,3,1/3) \in 50^*_S$ and $(6,1,-2/3) \in 15_S$. The latter is the only case with four external left-handed quarks. While both contributions can yield unsuppressed $n$--$\overline{n}$ transition rates we assume, for simplicity, that the former contribution dominates as it involves only right-handed SM fermions in the external legs. This particular choice of dominance does not affect the overall conclusions presented below.

Note that for all minimal, phenomenologically viable neutrino mass models, the neutrino mass and corresponding $n$--$\overline{n}$ diagrams both occur at the same perturbative order. Our discussion sets aside issues such as charged fermion mass generation and the viability of gauge coupling unification. A study addressing these aspects and their experimental implications will be presented elsewhere. We will, however, estimate whether a viable neutrino mass scale is compatible with the experimentally accessible $n$--$\overline{n}$ signal.

\begin{figure}[t!]
\centering
\includegraphics[width=0.45\textwidth]{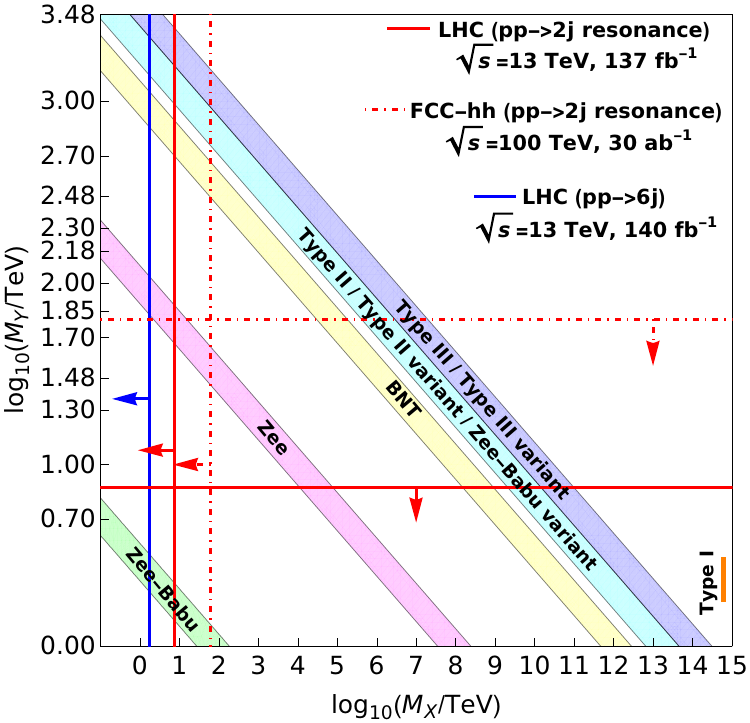}
\caption{$Y= S_1$ in all models. For BNT model, we  fix the mass of  $F_1$ to be equal to the mass of $S_2$. $X$ is to be inferred from Table~\ref{tab:EFT}. \texttt{LHC}/\texttt{FCC-hh} $pp\to 2j$~\cite{CMS:2019gwf,Harris:2022kls,Bernardi:2022hny} ($pp\to 6j$~\cite{ATLAS:2024kqk}) bound is applicable to di-quark scalars (color octet fermion). Each band corresponds to observable range of  $n$--$\overline{n}$ process with parameter space to the left ruled out by SK, whereas the region to the right is beyond experimental sensitivity. } \label{fig:NNBAR}
\end{figure}

The fields responsible for the neutrino mass generation are specified in Table~\ref{tab:EFT}.
To generate experimentally viable neutrino mass scale~\cite{Esteban:2024eli} one requires $\Lambda_1/|C^{\nu\nu}_{\tau\tau}| \simeq 6\times 10^{14}\,\mathrm{GeV}$, whereas current bounds from \texttt{KamLAND-Zen}~\cite{KamLAND-Zen:2024eml} and the projected reach of \texttt{nEXO}~\cite{nEXO:2021ujk} on $0\nu\beta\beta$ imply $6\times 10^{15}\,\mathrm{GeV} \gtrsim \Lambda_1/|C^{\nu\nu}_{ee}| \gtrsim 2.5\times 10^{14}\,\mathrm{GeV}$. We also list in Table~\ref{tab:EFT} the scales and associated couplings relevant for $n$--$\overline{n}$ oscillation, as introduced in Eq.\ \eqref{eq:d9operator}, for all models under consideration. The corresponding experimental accessibility of $n$--$\overline{n}$ oscillation signal is shown in Fig.~\ref{fig:NNBAR}, where we implement the current bound from \texttt{Super-Kamiokande}~\cite{Super-Kamiokande:2020bov} and future reach from \texttt{NNBAR}~\cite{Addazi:2020nlz}. We set all Yukawa couplings to one and the relevant cubic scalar coupling to $\widetilde{\mu} = M_{S_2}$. 

Fig.\ \ref{fig:NNBAR} demonstrates that if $n$--$\overline{n}$ oscillation signal is observed, the Zee-Babu and the Type I seesaw models cannot be its source due to \texttt{LHC} bounds on dijet resonances. Also, the Type I, II, III, and BNT models may be in tension with proton decay limits unless the leptoquark couplings of the color triplet are extremely suppressed. For example, if one is to have a TeV scale color triplet with order-one diquark couplings, its lepton–quark Yukawa couplings should be $\lesssim 10^{-24}$ to sufficiently suppress associated proton decay signatures~\cite{Dorsner:2012uz}. One possible approach is to abandon renormalization requirement altogether and consider higher-dimensional operators that allow for arbitrary suppression of the relevant color triplet Yukawa interactions, as discussed in Refs.\ \cite{Dorsner:2024seb,Dorsner:2025rzj}. The exact suppression implementation would still depend on the details of the model and is beyond the scope of this work. This is something one should keep in mind when interpreting the last column of Table~\ref{tab:EFT}. On the other hand, the variants of the Type II, III, and Zee-Babu models, as well as the Zee model itself are all viable with regard to proton decay constraints and may potentially yield collider-accessible signatures.

While Fig.~\ref{fig:NNBAR} displays the most optimistic scenario with regard to observation of $n$--$\overline{n}$ oscillation signal, we clarify how the picture would change if the relevant Yukawa or cubic couplings are varied. 

Let us first address the Type II scenario. The oscillation lifetime exhibits the following parameter dependence~\cite{Dorsner:2025epy}
\begin{align}
\tau_\mathrm{\bf{A}} = \frac{20}{3\mathcal M}\;\frac{1}{y^{dd}(y^{ud})^2}\frac{M_{S_1}^4 M_{S_2}^2}{\widetilde{\mu}},
\end{align}
where \textbf{A} denotes the topology type and $\mathcal M$ is the relevant hadronic matrix element~\cite{Buchoff:2015qwa,Rinaldi:2018osy,Rinaldi:2019thf,Fridell:2021gag,ThomasArun:2025rgx}. Consequently, achieving oscillation times within the observable range with smaller Yukawa coupling(s) would require lighter mediator state(s), implying that the observation bands shown in Fig.\ \ref{fig:NNBAR} would  shift diagonally downward. A similar effect would occur if the cubic coupling $\widetilde{\mu}$ is reduced. Note that this dimensionful coupling is  constrained from above, i.e., $\widetilde{\mu} \lesssim 3\times \mathrm{max}(M_{S_1}, M_{S_2})$, if one is to avoid charge-breaking minima~\cite{Frere:1983ag,Alvarez-Gaume:1983drc,Casas:1996de}. We, in particular, use $\widetilde{\mu} =M_{S_2}$ to generate Fig.\ \ref{fig:NNBAR}.

For the Type III seesaw, the parameter dependence is~\cite{Dorsner:2025epy}
\begin{align}
\tau_\mathrm{\textbf{B}} =\frac{1}{\mathcal M}\; \frac{1}{(y^{dd})^2 (y^{ud})^2} M_{F_1} M_{S_1}^4,
\end{align}
where \textbf{B} denotes the topology type. If one would reduce the relevant Yukawa couplings, the viable new physics mass scale would again need to go to lower values in order to have observable $n$--$\overline{n}$ oscillation signal.  For instance, assuming unity for all Yukawa couplings and degenerate masses $M_{F_1} \approx M_{S_1} \approx 450$\,TeV, the $n$--$\bar{n}$ transition rate is potentially measurable. If these couplings are reduced by an order of magnitude, the mass scales must decrease to $M_{F_1} \approx M_{S_1} \approx 70$\,TeV to maintain observability. The experimentally accessible range is bounded from below by the Super-Kamiokande limit $\tau_{n-\overline{n}}^\mathrm{SK} \geq 4.7\times 10^8\,\mathrm{s}$~\cite{Super-Kamiokande:2020bov}, while the upper bound is set by the  NNBAR experimental sensitivity
$\tau_{n-\overline{n}}^\mathrm{NNBAR} \leq 3\times 10^9\,\mathrm{s}$~\cite{Addazi:2020nlz}.

The case of the Type I seesaw is slightly different since the mass $M_{F_1}$ not only enters $n$--$\overline{n}$ prediction but it is also directly connected to the neutrino mass scale. For instance, if all Yukawa couplings are reduced by a factor of $10$, $M_{F_1}$ should need to be simultaneously lowered by a factor of $10^{2}$ in order to preserve the correct neutrino mass scale. With these changes, an observable $n$--$\overline{n}$ oscillation could still be achieved but only if $M_{S_1}$ is in the sub-TeV range. This, on the other hand, would be in conflict with the \texttt{LHC} data. The same issue arises if one is to reduce diquark couplings entering prediction for the $n$--$\overline{n}$ oscillation signal within the Type I seesaw scenario. In short, Fig.~\ref{fig:NNBAR} depicts expectations for all possible scenarios considered if the relevant Yukawa couplings are set to one.

In the present work, all the couplings and mass parameters are evaluated at the $n$--$\overline{n}$ mediator scale set by max($M_{S_{i}}$) and/or max($M_{F_{j}}$). We expect renormalization-group (RG) effects on the tree level and radiative neutrino-mass generation sectors to induce at most moderate quantitative corrections without qualitative impact on the viable parameter regions shown in Fig.~7. For example, since the mediator masses in the Zee-Babu model lie in the multi-TeV range, the logarithmic running between the mediator and electroweak scales should be relatively modest. Also, our analysis is intended as a leading-order phenomenological study rather than a precision calculation. A complete effective field theory matching and RG evolution analysis for the scenarios we address here would become a necessity if and when $n$--$\overline{n}$ oscillation signal is observed.

Moreover, in plotting Fig.~\ref{fig:NNBAR}, we adopt the relevant hadronic matrix elements from Table~1 of Ref.~\cite{Fridell:2021gag}. Although for the $d=9$ operators we do not explicitly include the RG running of the corresponding transition matrix elements, these effects are known to induce only moderate numerical corrections. In particular, Ref.~\cite{Fridell:2021gag} evolves the matrix elements from $\mu_0 = 2\,\mathrm{GeV}$ to the relevant new-physics scale (even up to the GUT scale) to find that they remain relatively stable under RG evolution, preserving the same order of magnitude across the phenomenologically relevant mass range. Even the largest RG-induced variation over the full range from the GUT scale down to low energies corresponds only to an enhancement of the matrix element by a factor of a few (in particular, for $\mathcal{M}_5$, it is a factor of three enhancement)~\cite{Fridell:2021gag}. Therefore, for the mediator mass scales considered in this work, we do not expect the omission of these running effects to qualitatively alter the viable parameter regions shown in Fig.~\ref{fig:NNBAR}.

\section{Conclusions}\label{sec:04}
We have demonstrated that Majorana neutrino mass generation and $n$--$\overline{n}$ oscillation are intrinsically linked if $SU(5)$ is realized in nature. Our study explores the simplest extensions of the GG model that yield realistic neutrino sector and, in doing so, automatically induce $n$--$\overline{n}$ oscillation. This implies the existence of additional baryon-number-violating processes beyond the ever-present proton decay mediated by gauge bosons and scalars. Majorana masses can be probed via $|\Delta L|=2$ processes, such as neutrinoless double beta decay and lepton flavor–violating transitions like muon to positron conversion. Similarly, $|\Delta B|=2$ processes, including $n$--$\overline{n}$ oscillation and di-nucleon decays, offer complementary experimental avenues, motivating dedicated searches for nucleon decay and neutron–antineutron oscillations.

\vspace{5pt}
We summarize our crucial findings here:
\begin{itemize}

\item Within the economical models considered in this work, the mediators responsible for neutron--antineutron oscillations can be either scalar or fermionic states. The scalar mediators include color-triplet and color-sextet scalars, while the beyond SM fermions can be singlets, color sextets, or color octets.

\item Models that contain at least one scalar color triplet $(3,1,-1/3)$ mediator cannot lead to observable $n$--$\overline{n}$ oscillations because of the associated proton decay constraints. Consequently, an observation of $n$--$\overline{n}$ oscillations in future experiments would strongly disfavor the minimal versions of Type I, Type II, and Type III seesaw models, as well as the BNT model.

\item The Zee-Babu model cannot generate observable $n$--$\overline{n}$ oscillations since it requires scalar diquarks --- a color triplet $(\overline{3},1,-2/3)$ and a color sextet --- in a mass range already excluded by \texttt{LHC} searches. This exclusion originates from the antisymmetric Yukawa couplings of the color-triplet scalar, where two such mediators are needed to generate $n$--$\overline{n}$ process.

\item Models compatible with observable $n$--$\overline{n}$ oscillations include the minimally extended Type II and Type III seesaw models, the extended Zee-Babu model, and the minimal Zee model. In the Type II and Zee-Babu scenarios, the relevant mediators are scalar color sextets that do not induce proton decay. In the Zee model, one mediator is the color-triplet scalar $(\overline{3},1,-2/3)$, which is likewise free from proton decay constraints. In the extended Type III scenario, the oscillation is mediated by a color-sextet scalar together with a color-octet fermion, again avoiding proton decay bounds.

\item If future experiments observe $n$--$\overline{n}$ oscillations, the Zee model discussed here will be fully testable at the \texttt{FCC-hh} through searches for low-mass diquarks. In contrast, even in the absence of such a discovery at the \texttt{FCC-hh}, viable parameter space for observable $n$--$\overline{n}$ oscillation would still remain for the Type II and Type III seesaw model variants, as well as the Zee-Babu model variant.

\end{itemize}

Our dedicated study, therefore, provides clear guidance for model building in the exciting event of an observed $n$--$\overline{n}$ oscillation signal.

\vspace{10pt}
\section*{Acknowledgments}
I.D.\ acknowledges the financial support from the Slovenian Research Agency (research core funding No.\ P1-0035) and hospitality of the CERN Theory Department, where part of this work was completed. S.F.\ and S.S.\ acknowledge the financial support
from the Slovenian Research Agency (research core funding No.\ P1-0035 and N1-0321). 

\bibliographystyle{style}
\bibliography{reference}

\providecommand{\href}[2]{#2}\begingroup\raggedright\begin{thebibliography}{10}

\bibitem{Georgi:1974sy}
H.~Georgi and S.~L. Glashow, ``{Unity of All Elementary Particle Forces},''
\href{http://dx.doi.org/10.1103/PhysRevLett.32.438}{{\em Phys. Rev. Lett.} {\bfseries 32} (1974) 438--441}.

\bibitem{Dorsner:2025epy}
I.~Dorsner, S.~Fajfer, and S.~Saad, ``{Beyond Neutrino Mass: Observable $n$-$\overline{n}$ Oscillations in UV Complete Seesaw Models},'' \href{http://arxiv.org/abs/2509.00145}{{\ttfamily arXiv:2509.00145 [hep-ph]}}.

\bibitem{Kuzmin:1970nx}
V.~Kuzmin, ``{Cp violation and baryon asymmetry of the universe},'' {\em Pisma Zh. Eksp. Teor. Fiz.} {\bfseries 12} (1970) 335--337.

\bibitem{Mohapatra:1980qe}
R.~N. Mohapatra and R.~Marshak, ``{Local B-L Symmetry of Electroweak Interactions, Majorana Neutrinos and Neutron Oscillations},'' \href{http://dx.doi.org/10.1103/PhysRevLett.44.1316}{{\em Phys. Rev. Lett.} {\bfseries 44} (1980) 1316--1319}. [Erratum: Phys.Rev.Lett. 44, 1643 (1980)].

\bibitem{Phillips:2014fgb}
I.~Phillips, D.G. {\em et~al.}, ``{Neutron-Antineutron Oscillations: Theoretical Status and Experimental Prospects},'' \href{http://dx.doi.org/10.1016/j.physrep.2015.11.001}{{\em Phys. Rept.} {\bfseries 612} (2016) 1--45}, \href{http://arxiv.org/abs/1410.1100}{{\ttfamily arXiv:1410.1100 [hep-ex]}}.

\bibitem{Addazi:2020nlz}
A.~Addazi {\em et~al.}, ``{New high-sensitivity searches for neutrons converting into antineutrons and/or sterile neutrons at the HIBEAM/NNBAR experiment at the European Spallation Source},'' \href{http://dx.doi.org/10.1088/1361-6471/abf429}{{\em J. Phys. G} {\bfseries 48} no.~7, (2021) 070501}, \href{http://arxiv.org/abs/2006.04907}{{\ttfamily arXiv:2006.04907 [physics.ins-det]}}.

\bibitem{Rao:1983sd}
S.~Rao and R.~E. Shrock, ``{Six Fermion ($B-L$) Violating Operators of Arbitrary Generational Structure},'' \href{http://dx.doi.org/10.1016/0550-3213(84)90365-1}{{\em Nucl. Phys. B} {\bfseries 232} (1984) 143--179}.

\bibitem{Dolinski:2019nrj}
M.~J. Dolinski, A.~W.~P. Poon, and W.~Rodejohann, ``{Neutrinoless Double-Beta Decay: Status and Prospects},'' \href{http://dx.doi.org/10.1146/annurev-nucl-101918-023407}{{\em Ann. Rev. Nucl. Part. Sci.} {\bfseries 69} (2019) 219--251}, \href{http://arxiv.org/abs/1902.04097}{{\ttfamily arXiv:1902.04097 [nucl-ex]}}.

\bibitem{Super-Kamiokande:2020bov}
{\bfseries Super-Kamiokande} Collaboration, K.~Abe {\em et~al.}, ``{Neutron-antineutron oscillation search using a 0.37 megaton-years exposure of Super-Kamiokande},'' \href{http://dx.doi.org/10.1103/PhysRevD.103.012008}{{\em Phys. Rev. D} {\bfseries 103} no.~1, (2021) 012008}, \href{http://arxiv.org/abs/2012.02607}{{\ttfamily arXiv:2012.02607 [hep-ex]}}.

\bibitem{DUNE:2015lol}
{\bfseries DUNE} Collaboration, R.~Acciarri {\em et~al.}, ``{Long-Baseline Neutrino Facility (LBNF) and Deep Underground Neutrino Experiment (DUNE)}: {Conceptual Design Report, Volume 2: The Physics Program for DUNE at LBNF},'' \href{http://arxiv.org/abs/1512.06148}{{\ttfamily arXiv:1512.06148 [physics.ins-det]}}.

\bibitem{Lee:2021hnx}
M.~Lee and M.~MacKenzie, ``{Muon to Positron Conversion},'' \href{http://dx.doi.org/10.3390/universe8040227}{{\em Universe} {\bfseries 8} (2022) 227}, \href{http://arxiv.org/abs/2110.07093}{{\ttfamily arXiv:2110.07093 [hep-ex]}}.

\bibitem{Aitken:2017wie}
K.~Aitken, D.~McKeen, T.~Neder, and A.~E. Nelson, ``{Baryogenesis from Oscillations of Charmed or Beautiful Baryons},'' \href{http://dx.doi.org/10.1103/PhysRevD.96.075009}{{\em Phys. Rev. D} {\bfseries 96} no.~7, (2017) 075009}, \href{http://arxiv.org/abs/1708.01259}{{\ttfamily arXiv:1708.01259 [hep-ph]}}.

\bibitem{Girmohanta:2019cjm}
S.~Girmohanta and R.~Shrock, ``{Improved Upper Limits on Baryon-Number Violating Dinucleon Decays to Dileptons},'' \href{http://dx.doi.org/10.1016/j.physletb.2020.135296}{{\em Phys. Lett. B} {\bfseries 803} (2020) 135296}, \href{http://arxiv.org/abs/1910.08356}{{\ttfamily arXiv:1910.08356 [hep-ph]}}.

\bibitem{Dev:2022jbf}
P.~S.~B. Dev {\em et~al.}, ``{Searches for baryon number violation in neutrino experiments: a white paper},'' \href{http://dx.doi.org/10.1088/1361-6471/ad1658}{{\em J. Phys. G} {\bfseries 51} no.~3, (2024) 033001}, \href{http://arxiv.org/abs/2203.08771}{{\ttfamily arXiv:2203.08771 [hep-ex]}}.

\bibitem{Beneito:2025ond}
A.~B. Beneito, I, S.~Fajfer, and A.~A. Petrov, ``{New Avenues for $|\Delta B|$ = 2 Processes Beyond Neutron-Antineutron Oscillations},'' \href{http://arxiv.org/abs/2511.05657}{{\ttfamily arXiv:2511.05657 [hep-ph]}}.

\bibitem{Weinberg:1979sa}
S.~Weinberg, ``{Baryon and Lepton Nonconserving Processes},''
\href{http://dx.doi.org/10.1103/PhysRevLett.43.1566}{{\em Phys. Rev. Lett.} {\bfseries 43} (1979) 1566--1570}.

\bibitem{Weinberg:1980bf}
S.~Weinberg, ``{Varieties of Baryon and Lepton Nonconservation},'' \href{http://dx.doi.org/10.1103/PhysRevD.22.1694}{{\em Phys. Rev. D} {\bfseries 22} (1980) 1694}.

\bibitem{Wilczek:1979et}
F.~Wilczek and A.~Zee, ``{Conservation or Violation of $B^-$l in Proton Decay},'' \href{http://dx.doi.org/10.1016/0370-2693(79)90475-1}{{\em Phys. Lett. B} {\bfseries 88} (1979) 311--314}.

\bibitem{Wilczek:1979hc}
F.~Wilczek and A.~Zee, ``{Operator Analysis of Nucleon Decay},'' \href{http://dx.doi.org/10.1103/PhysRevLett.43.1571}{{\em Phys. Rev. Lett.} {\bfseries 43} (1979) 1571--1573}.

\bibitem{Weldon:1980gi}
H.~A. Weldon and A.~Zee, ``{Operator Analysis of New Physics},'' \href{http://dx.doi.org/10.1016/0550-3213(80)90218-7}{{\em Nucl. Phys. B} {\bfseries 173} (1980) 269--290}.

\bibitem{Georgi:1979df}
H.~Georgi and C.~Jarlskog, ``{A New Lepton - Quark Mass Relation in a Unified Theory},'' \href{http://dx.doi.org/10.1016/0370-2693(79)90842-6}{{\em Phys. Lett. B} {\bfseries 86} (1979) 297--300}.

\bibitem{deGouvea:2014lva}
A.~de~Gouvea, J.~Herrero-Garcia, and A.~Kobach, ``{Neutrino Masses, Grand Unification, and Baryon Number Violation},'' \href{http://dx.doi.org/10.1103/PhysRevD.90.016011}{{\em Phys. Rev. D} {\bfseries 90} no.~1, (2014) 016011}, \href{http://arxiv.org/abs/1404.4057}{{\ttfamily arXiv:1404.4057 [hep-ph]}}.

\bibitem{Minkowski:1977sc}
P.~Minkowski, ``{$\mu \to e\gamma$ at a Rate of One Out of $10^{9}$ Muon Decays?},''
\href{http://dx.doi.org/10.1016/0370-2693(77)90435-X}{{\em Phys. Lett.} {\bfseries 67B} (1977) 421--428}.

\bibitem{Yanagida:1979as}
T.~Yanagida, ``{Horizontal gauge symmetry and masses of neutrinos},''
{\em Conf. Proc.} {\bfseries C7902131} (1979) 95--99.

\bibitem{Glashow:1979nm}
S.~Glashow, ``{The Future of Elementary Particle Physics},'' \href{http://dx.doi.org/10.1007/978-1-4684-7197-7\_15}{{\em NATO Sci. Ser. B} {\bfseries 61} (1980) 687}.

\bibitem{Gell-Mann:1979vob}
M.~Gell-Mann, P.~Ramond, and R.~Slansky, ``{Complex Spinors and Unified Theories},'' {\em Conf. Proc. C} {\bfseries 790927} (1979) 315--321, \href{http://arxiv.org/abs/1306.4669}{{\ttfamily arXiv:1306.4669 [hep-th]}}.

\bibitem{Mohapatra:1979ia}
R.~N. Mohapatra and G.~Senjanovic, ``{Neutrino Mass and Spontaneous Parity Nonconservation},'' \href{http://dx.doi.org/10.1103/PhysRevLett.44.912}{{\em Phys. Rev. Lett.} {\bfseries 44} (1980) 912}.

\bibitem{Magg:1980ut}
M.~Magg and C.~Wetterich, ``{Neutrino Mass Problem and Gauge Hierarchy},'' \href{http://dx.doi.org/10.1016/0370-2693(80)90825-4}{{\em Phys. Lett. B} {\bfseries 94} (1980) 61--64}.

\bibitem{Schechter:1980gr}
J.~Schechter and J.~W.~F. Valle, ``{Neutrino Masses in SU(2) x U(1) Theories},''
\href{http://dx.doi.org/10.1103/PhysRevD.22.2227}{{\em Phys. Rev.} {\bfseries D22} (1980) 2227}.

\bibitem{Lazarides:1980nt}
G.~Lazarides, Q.~Shafi, and C.~Wetterich, ``{Proton Lifetime and Fermion Masses in an SO(10) Model},'' \href{http://dx.doi.org/10.1016/0550-3213(81)90354-0}{{\em Nucl. Phys. B} {\bfseries 181} (1981) 287--300}.

\bibitem{Mohapatra:1980yp}
R.~N. Mohapatra and G.~Senjanovic, ``{Neutrino Masses and Mixings in Gauge Models with Spontaneous Parity Violation},'' \href{http://dx.doi.org/10.1103/PhysRevD.23.165}{{\em Phys. Rev. D} {\bfseries 23} (1981) 165}.

\bibitem{Ma:1998dn}
E.~Ma, ``{Pathways to naturally small neutrino masses},'' \href{http://dx.doi.org/10.1103/PhysRevLett.81.1171}{{\em Phys. Rev. Lett.} {\bfseries 81} (1998) 1171--1174}, \href{http://arxiv.org/abs/hep-ph/9805219}{{\ttfamily arXiv:hep-ph/9805219}}.

\bibitem{Dorsner:2005fq}
I.~Dorsner and P.~Fileviez~Perez, ``{Unification without supersymmetry: Neutrino mass, proton decay and light leptoquarks},'' \href{http://dx.doi.org/10.1016/j.nuclphysb.2005.06.016}{{\em Nucl. Phys.} {\bfseries B723} (2005) 53--76},
\href{http://arxiv.org/abs/hep-ph/0504276}{{\ttfamily arXiv:hep-ph/0504276 [hep-ph]}}.

\bibitem{Dorsner:2005ii}
I.~Dorsner, P.~Fileviez~Perez, and R.~Gonzalez~Felipe, ``{Phenomenological and cosmological aspects of a minimal GUT scenario},'' \href{http://dx.doi.org/10.1016/j.nuclphysb.2006.05.006}{{\em Nucl. Phys. B} {\bfseries 747} (2006) 312--327}, \href{http://arxiv.org/abs/hep-ph/0512068}{{\ttfamily arXiv:hep-ph/0512068}}.

\bibitem{Dorsner:2006hw}
I.~Dorsner, P.~Fileviez~Perez, and G.~Rodrigo, ``{Fermion masses and the UV cutoff of the minimal realistic SU(5)},'' \href{http://dx.doi.org/10.1103/PhysRevD.75.125007}{{\em Phys. Rev. D} {\bfseries 75} (2007) 125007}, \href{http://arxiv.org/abs/hep-ph/0607208}{{\ttfamily arXiv:hep-ph/0607208}}.

\bibitem{Dorsner:2007fy}
I.~Dorsner and I.~Mocioiu, ``{Predictions from type II see-saw mechanism in SU(5)},'' \href{http://dx.doi.org/10.1016/j.nuclphysb.2007.12.004}{{\em Nucl. Phys. B} {\bfseries 796} (2008) 123--136}, \href{http://arxiv.org/abs/0708.3332}{{\ttfamily arXiv:0708.3332 [hep-ph]}}.

\bibitem{Antusch:2022afk}
S.~Antusch, K.~Hinze, and S.~Saad, ``{Viable quark-lepton Yukawa ratios and nucleon decay predictions in SU(5) GUTs with type-II seesaw},'' \href{http://dx.doi.org/10.1016/j.nuclphysb.2022.116049}{{\em Nucl. Phys. B} {\bfseries 986} (2023) 116049}, \href{http://arxiv.org/abs/2205.01120}{{\ttfamily arXiv:2205.01120 [hep-ph]}}.

\bibitem{Calibbi:2022wko}
L.~Calibbi and X.~Gao, ``{Lepton flavor violation in minimal grand unified type II seesaw models},'' \href{http://dx.doi.org/10.1103/PhysRevD.106.095036}{{\em Phys. Rev. D} {\bfseries 106} no.~9, (2022) 095036}, \href{http://arxiv.org/abs/2206.10682}{{\ttfamily arXiv:2206.10682 [hep-ph]}}.

\bibitem{Antusch:2023mqe}
S.~Antusch, K.~Hinze, and S.~Saad, ``{Minimal SU(5) GUTs with vectorlike fermions},'' \href{http://dx.doi.org/10.1103/PhysRevD.108.095010}{{\em Phys. Rev. D} {\bfseries 108} no.~9, (2023) 095010}, \href{http://arxiv.org/abs/2308.08585}{{\ttfamily arXiv:2308.08585 [hep-ph]}}.

\bibitem{Kaladharan:2024bop}
A.~Kaladharan and S.~Saad, ``{Unified origin of inflation, baryon asymmetry, and neutrino mass},'' \href{http://dx.doi.org/10.1103/PhysRevD.110.116012}{{\em Phys. Rev. D} {\bfseries 110} no.~11, (2024) 116012}, \href{http://arxiv.org/abs/2409.02225}{{\ttfamily arXiv:2409.02225 [hep-ph]}}.

\bibitem{Dev:2025sox}
P.~S.~B. Dev, S.~Goswami, D.~Pachhar, and S.~K. Shukla, ``{Scalar-induced Neutrinoless Double Beta Decay in $SU(5)$},'' \href{http://arxiv.org/abs/2507.16606}{{\ttfamily arXiv:2507.16606 [hep-ph]}}.

\bibitem{Foot:1988aq}
R.~Foot, H.~Lew, X.~G. He, and G.~C. Joshi, ``{Seesaw Neutrino Masses Induced by a Triplet of Leptons},'' \href{http://dx.doi.org/10.1007/BF01415558}{{\em Z. Phys. C} {\bfseries 44} (1989) 441}.

\bibitem{Bajc:2006ia}
B.~Bajc and G.~Senjanovic, ``{Seesaw at LHC},'' \href{http://dx.doi.org/10.1088/1126-6708/2007/08/014}{{\em JHEP} {\bfseries 08} (2007) 014},
\href{http://arxiv.org/abs/hep-ph/0612029}{{\ttfamily arXiv:hep-ph/0612029 [hep-ph]}}.

\bibitem{Dorsner:2006fx}
I.~Dorsner and P.~Fileviez~Perez, ``{Upper Bound on the Mass of the Type III Seesaw Triplet in an SU(5) Model},'' \href{http://dx.doi.org/10.1088/1126-6708/2007/06/029}{{\em JHEP} {\bfseries 06} (2007) 029},
\href{http://arxiv.org/abs/hep-ph/0612216}{{\ttfamily arXiv:hep-ph/0612216 [hep-ph]}}.

\bibitem{FileviezPerez:2007bcw}
P.~Fileviez~Perez, ``{Renormalizable adjoint SU(5)},'' \href{http://dx.doi.org/10.1016/j.physletb.2007.07.075}{{\em Phys. Lett. B} {\bfseries 654} (2007) 189--193}, \href{http://arxiv.org/abs/hep-ph/0702287}{{\ttfamily arXiv:hep-ph/0702287}}.

\bibitem{Antusch:2023kli}
S.~Antusch, K.~Hinze, and S.~Saad, ``{Quark-lepton Yukawa ratios and nucleon decay in SU(5) GUTs with type-III seesaw},'' \href{http://dx.doi.org/10.1016/j.nuclphysb.2023.116195}{{\em Nucl. Phys. B} {\bfseries 991} (2023) 116195}, \href{http://arxiv.org/abs/2301.03601}{{\ttfamily arXiv:2301.03601 [hep-ph]}}.

\bibitem{Babu:2009aq}
K.~S. Babu, S.~Nandi, and Z.~Tavartkiladze, ``{New Mechanism for Neutrino Mass Generation and Triply Charged Higgs Bosons at the LHC},'' \href{http://dx.doi.org/10.1103/PhysRevD.80.071702}{{\em Phys. Rev.} {\bfseries D80} (2009) 071702},
\href{http://arxiv.org/abs/0905.2710}{{\ttfamily arXiv:0905.2710 [hep-ph]}}.

\bibitem{Dorsner:2019vgf}
I.~Dor\v{s}ner and S.~Saad, ``{Towards Minimal $SU(5)$},'' \href{http://dx.doi.org/10.1103/PhysRevD.101.015009}{{\em Phys. Rev. D} {\bfseries 101} no.~1, (2020) 015009}, \href{http://arxiv.org/abs/1910.09008}{{\ttfamily arXiv:1910.09008 [hep-ph]}}.

\bibitem{Dorsner:2021qwg}
I.~Dor\v{s}ner, E.~D\v{z}aferovi\'c-Ma\v{s}i\'c, and S.~Saad, ``{Parameter space exploration of the minimal SU(5) unification},'' \href{http://dx.doi.org/10.1103/PhysRevD.104.015023}{{\em Phys. Rev. D} {\bfseries 104} no.~1, (2021) 015023}, \href{http://arxiv.org/abs/2105.01678}{{\ttfamily arXiv:2105.01678 [hep-ph]}}.

\bibitem{Antusch:2023jok}
S.~Antusch, I.~Dor\v{s}ner, K.~Hinze, and S.~Saad, ``{Fully testable axion dark matter within a minimal SU(5) GUT},'' \href{http://dx.doi.org/10.1103/PhysRevD.108.015025}{{\em Phys. Rev. D} {\bfseries 108} no.~1, (2023) 015025}, \href{http://arxiv.org/abs/2301.00809}{{\ttfamily arXiv:2301.00809 [hep-ph]}}.

\bibitem{Dorsner:2024jiy}
I.~Dor\v{s}ner, E.~D\v{z}aferovi\'c-Ma\v{s}i\'c, S.~Fajfer, and S.~Saad, ``{Gauge and scalar boson mediated proton decay in a predictive SU(5) GUT model},'' \href{http://dx.doi.org/10.1103/PhysRevD.109.075023}{{\em Phys. Rev. D} {\bfseries 109} no.~7, (2024) 075023}, \href{http://arxiv.org/abs/2401.16907}{{\ttfamily arXiv:2401.16907 [hep-ph]}}.

\bibitem{Zee:1980ai}
A.~Zee, ``{A Theory of Lepton Number Violation, Neutrino Majorana Mass, and Oscillation},'' \href{http://dx.doi.org/10.1016/0370-2693(80)90349-4}{{\em Phys. Lett. B} {\bfseries 93} (1980) 389}. [Erratum: Phys.Lett.B 95, 461 (1980)].

\bibitem{Perez:2016qbo}
P.~Fileviez~Perez and C.~Murgui, ``{Renormalizable SU(5) Unification},'' \href{http://dx.doi.org/10.1103/PhysRevD.94.075014}{{\em Phys. Rev.} {\bfseries D94} no.~7, (2016) 075014},
\href{http://arxiv.org/abs/1604.03377}{{\ttfamily arXiv:1604.03377 [hep-ph]}}.

\bibitem{Klein:2019jgb}
C.~Klein, M.~Lindner, and S.~Vogl, ``{Radiative neutrino masses and successful $SU(5)$ unification},'' \href{http://dx.doi.org/10.1103/PhysRevD.100.075024}{{\em Phys. Rev.} {\bfseries D100} no.~7, (2019) 075024},
\href{http://arxiv.org/abs/1907.05328}{{\ttfamily arXiv:1907.05328 [hep-ph]}}.

\bibitem{Dorsner:2025rzj}
I.~Dor{\v{s}}ner, M.~Matkovi{\'c}, and S.~Saad, ``{Nonrenormalizable SU(5) GUTs: Leptoquark-induced neutrino masses},'' \href{http://dx.doi.org/10.1103/3plx-vwrb}{{\em Phys. Rev. D} {\bfseries 111} no.~11, (2025) 115039}, \href{http://arxiv.org/abs/2504.16022}{{\ttfamily arXiv:2504.16022 [hep-ph]}}.

\bibitem{Cheng:1980qt}
T.~P. Cheng and L.-F. Li, ``{Neutrino Masses, Mixings and Oscillations in SU(2) x U(1) Models of Electroweak Interactions},'' \href{http://dx.doi.org/10.1103/PhysRevD.22.2860}{{\em Phys. Rev. D} {\bfseries 22} (1980) 2860}.

\bibitem{Babu:1988ki}
K.~S. Babu, ``{Model of 'Calculable' Majorana Neutrino Masses},'' \href{http://dx.doi.org/10.1016/0370-2693(88)91584-5}{{\em Phys. Lett. B} {\bfseries 203} (1988) 132--136}.

\bibitem{Babu:2024jdw}
K.~S. Babu and S.~Saad, ``{Ultraviolet completion of a two-loop neutrino mass model},'' \href{http://dx.doi.org/10.1007/JHEP03(2025)132}{{\em JHEP} {\bfseries 03} (2025) 132}, \href{http://arxiv.org/abs/2412.14562}{{\ttfamily arXiv:2412.14562 [hep-ph]}}.

\bibitem{Saad:2019vjo}
S.~Saad, ``{Origin of a two-loop neutrino mass from SU(5) grand unification},'' \href{http://dx.doi.org/10.1103/PhysRevD.99.115016}{{\em Phys. Rev.} {\bfseries D99} no.~11, (2019) 115016},
\href{http://arxiv.org/abs/1902.11254}{{\ttfamily arXiv:1902.11254 [hep-ph]}}.

\bibitem{CMS:2019gwf}
{\bfseries CMS} Collaboration, A.~M. Sirunyan {\em et~al.}, ``{Search for high mass dijet resonances with a new background prediction method in proton-proton collisions at $\sqrt{s} =$ 13 TeV},'' \href{http://dx.doi.org/10.1007/JHEP05(2020)033}{{\em JHEP} {\bfseries 05} (2020) 033}, \href{http://arxiv.org/abs/1911.03947}{{\ttfamily arXiv:1911.03947 [hep-ex]}}.

\bibitem{Harris:2022kls}
R.~M. Harris, E.~G. Guler, and Y.~Guler, ``{Sensitivity to Dijet Resonances at Proton-Proton Colliders},'' in {\em {Snowmass 2021}}.
\newblock 2, 2022.
\newblock \href{http://arxiv.org/abs/2202.03389}{{\ttfamily arXiv:2202.03389 [hep-ex]}}.

\bibitem{Bernardi:2022hny}
G.~Bernardi {\em et~al.}, ``{The Future Circular Collider: a Summary for the US 2021 Snowmass Process},'' \href{http://arxiv.org/abs/2203.06520}{{\ttfamily arXiv:2203.06520 [hep-ex]}}.

\bibitem{ATLAS:2024kqk}
{\bfseries ATLAS} Collaboration, G.~Aad {\em et~al.}, ``{A search for R-parity-violating supersymmetry in final states containing many jets in pp collisions at $ \sqrt{s} $ = 13 TeV with the ATLAS detector},'' \href{http://dx.doi.org/10.1007/JHEP05(2024)003}{{\em JHEP} {\bfseries 05} (2024) 003}, \href{http://arxiv.org/abs/2401.16333}{{\ttfamily arXiv:2401.16333 [hep-ex]}}.

\bibitem{Esteban:2024eli}
I.~Esteban, M.~C. Gonzalez-Garcia, M.~Maltoni, I.~Martinez-Soler, J.~a.~P. Pinheiro, and T.~Schwetz, ``{NuFit-6.0: updated global analysis of three-flavor neutrino oscillations},'' \href{http://dx.doi.org/10.1007/JHEP12(2024)216}{{\em JHEP} {\bfseries 12} (2024) 216}, \href{http://arxiv.org/abs/2410.05380}{{\ttfamily arXiv:2410.05380 [hep-ph]}}.

\bibitem{KamLAND-Zen:2024eml}
{\bfseries KamLAND-Zen} Collaboration, S.~Abe {\em et~al.}, ``{Search for Majorana Neutrinos with the Complete KamLAND-Zen Dataset},'' \href{http://arxiv.org/abs/2406.11438}{{\ttfamily arXiv:2406.11438 [hep-ex]}}.

\bibitem{nEXO:2021ujk}
{\bfseries nEXO} Collaboration, G.~Adhikari {\em et~al.}, ``{nEXO: neutrinoless double beta decay search beyond 10$^{28}$ year half-life sensitivity},'' \href{http://dx.doi.org/10.1088/1361-6471/ac3631}{{\em J. Phys. G} {\bfseries 49} no.~1, (2022) 015104}, \href{http://arxiv.org/abs/2106.16243}{{\ttfamily arXiv:2106.16243 [nucl-ex]}}.

\bibitem{Dorsner:2012uz}
I.~Dorsner, ``{A scalar leptoquark in SU(5)},'' \href{http://dx.doi.org/10.1103/PhysRevD.86.055009}{{\em Phys. Rev.} {\bfseries D86} (2012) 055009},
\href{http://arxiv.org/abs/1206.5998}{{\ttfamily arXiv:1206.5998 [hep-ph]}}.

\bibitem{Dorsner:2024seb}
I.~Dor{\v{s}}ner and S.~Saad, ``{Is doublet-triplet splitting necessary?},'' \href{http://dx.doi.org/10.1103/PhysRevD.110.075025}{{\em Phys. Rev. D} {\bfseries 110} no.~7, (2024) 075025}, \href{http://arxiv.org/abs/2404.09021}{{\ttfamily arXiv:2404.09021 [hep-ph]}}.

\bibitem{Buchoff:2015qwa}
M.~I. Buchoff and M.~Wagman, ``{Perturbative Renormalization of Neutron-Antineutron Operators},'' \href{http://dx.doi.org/10.1103/PhysRevD.93.016005}{{\em Phys. Rev. D} {\bfseries 93} no.~1, (2016) 016005}, \href{http://arxiv.org/abs/1506.00647}{{\ttfamily arXiv:1506.00647 [hep-ph]}}. [Erratum: Phys.Rev.D 98, 079901 (2018)].

\bibitem{Rinaldi:2018osy}
E.~Rinaldi, S.~Syritsyn, M.~L. Wagman, M.~I. Buchoff, C.~Schroeder, and J.~Wasem, ``{Neutron-antineutron oscillations from lattice QCD},'' \href{http://dx.doi.org/10.1103/PhysRevLett.122.162001}{{\em Phys. Rev. Lett.} {\bfseries 122} no.~16, (2019) 162001}, \href{http://arxiv.org/abs/1809.00246}{{\ttfamily arXiv:1809.00246 [hep-lat]}}.

\bibitem{Rinaldi:2019thf}
E.~Rinaldi, S.~Syritsyn, M.~L. Wagman, M.~I. Buchoff, C.~Schroeder, and J.~Wasem, ``{Lattice QCD determination of neutron-antineutron matrix elements with physical quark masses},'' \href{http://dx.doi.org/10.1103/PhysRevD.99.074510}{{\em Phys. Rev. D} {\bfseries 99} no.~7, (2019) 074510}, \href{http://arxiv.org/abs/1901.07519}{{\ttfamily arXiv:1901.07519 [hep-lat]}}.

\bibitem{Fridell:2021gag}
K.~Fridell, J.~Harz, and C.~Hati, ``{Probing baryogenesis with neutron-antineutron oscillations},'' \href{http://dx.doi.org/10.1007/JHEP11(2021)185}{{\em JHEP} {\bfseries 11} (2021) 185}, \href{http://arxiv.org/abs/2105.06487}{{\ttfamily arXiv:2105.06487 [hep-ph]}}.

\bibitem{ThomasArun:2025rgx}
M.~Thomas~Arun, S.~M, and R.~Pal, ``{RG evolution and effect of intermediate new physics on {\ensuremath{\Delta}}B = 2 six-quark operators},'' \href{http://dx.doi.org/10.1007/JHEP10(2025)032}{{\em JHEP} {\bfseries 10} (2025) 032}, \href{http://arxiv.org/abs/2506.10105}{{\ttfamily arXiv:2506.10105 [hep-ph]}}.

\bibitem{Frere:1983ag}
J.~M. Frere, D.~R.~T. Jones, and S.~Raby, ``{Fermion Masses and Induction of the Weak Scale by Supergravity},'' \href{http://dx.doi.org/10.1016/0550-3213(83)90606-5}{{\em Nucl. Phys. B} {\bfseries 222} (1983) 11--19}.

\bibitem{Alvarez-Gaume:1983drc}
L.~Alvarez-Gaume, J.~Polchinski, and M.~B. Wise, ``{Minimal Low-Energy Supergravity},'' \href{http://dx.doi.org/10.1016/0550-3213(83)90591-6}{{\em Nucl. Phys. B} {\bfseries 221} (1983) 495}.

\bibitem{Casas:1996de}
J.~A. Casas and S.~Dimopoulos, ``{Stability bounds on flavor violating trilinear soft terms in the MSSM},'' \href{http://dx.doi.org/10.1016/0370-2693(96)01000-3}{{\em Phys. Lett. B} {\bfseries 387} (1996) 107--112}, \href{http://arxiv.org/abs/hep-ph/9606237}{{\ttfamily arXiv:hep-ph/9606237}}.

\end{thebibliography}\endgroup
\end{document}